\documentclass[12pt]{article}
\usepackage{graphicx}
\usepackage{amsmath}
\usepackage{amssymb}
\usepackage{subfigure}
\usepackage{latexsym}
\usepackage{hyperref}
\usepackage[sort]{cite}

\def\Ket#1{\mathinner{|{#1}\rangle}}

\def\bra#1{\left<#1\right|}
\def\ket#1{\left|#1\right>}
{\catcode`\|=\active 
  \gdef\Braket#1{\left<\mathcode`\|"8000\let|\bravert {#1}\right>}}
\def\bravert{\egroup\,\vrule\,\bgroup}
\newcommand{\ext}{-\!\!\!-\!\!\!-\!\!\!-\!\!\!-\!\!\!-\!\!\!}
\newcommand{\alg}[1]{\mathfrak{#1}}

\newcommand{\su}{\alg{su}}
\newcommand{\psu}{\alg{psu}}
\newcommand{\Sl}{\alg{sl}}
\newcommand{\so}{\alg{so}}

\newcommand{\tr}{\mathop{\rm tr}}

\newcommand{\be}{\begin{eqnarray}}
\newcommand{\ee}{\end{eqnarray}}
\newcommand{\bc}{\begin{center}}
\newcommand{\ec}{\end{center}}
\newcommand{\bea}{\begin{eqnarray}}
\newcommand{\eea}{\end{eqnarray}}
\newcommand{\ben}{\begin{equation}}
\newcommand{\een}{\end{equation}}

\newcommand{\Rhat}{\widehat R}
\newcommand{\del}{\partial}

\newcommand{\nn}{\nonumber}

\numberwithin{equation}{section}

\begin{document}

\begin{titlepage}
\begin{flushright}
CALT-68-2546\\
hep-th/0504203
\end{flushright}
\vspace{15 mm}
\begin{center}
{\huge  Open string integrability and AdS/CFT  }
\end{center}
\vspace{12 mm}

\begin{center}
{\large
Tristan McLoughlin and
Ian Swanson }\\
\vspace{3mm}
California Institute of Technology\\
Pasadena, CA 91125, USA
\end{center}
\vspace{5 mm}
\begin{center}
{\large Abstract}
\end{center}
\noindent
We present a set of long-range Bethe ansatz equations for open quantum strings on
$AdS_5\times S^5$ and demonstrate that they diagonalize bosonic $\su(2)$ and
$\Sl(2)$ sectors of the theory in the near-pp-wave limit.  
Results are compared with energy spectra obtained by direct quantization
of the open string theory, and we find agreement in this limit to all orders
in the 't~Hooft coupling $\lambda = g_{\rm YM}^2 N_c$.  
We also propose long-range Bethe ans\"atze for $\su(2)$ and $\Sl(2)$ sectors of 
the dual ${\cal N}=2$ defect conformal field theory.
In accordance with previous investigations, we find exact agreement between 
string theory and gauge theory at one- and two-loop order in $\lambda$, but a 
general disagreement at higher loops.  It has been conjectured that the 
sudden mismatch at three-loop order may be due to long-range interaction terms 
that are lost in the weak coupling expansion of the gauge theory dilatation operator.
These terms are thought to include interactions that wrap around gauge-traced
operators, or around the closed-string worldsheet.
We comment on the role these interactions play in both open and closed sectors of
string states and operators near the pp-wave/BMN limit of the AdS/CFT correspondence.
\vspace{1cm}
\begin{flushleft}
\today
\end{flushleft}
\end{titlepage}
\newpage
\section{Introduction}
In 2002 Berenstein, Maldacena and Nastase demonstrated that the energy 
spectrum of type IIB string theory on a pp-wave background geometry 
can be matched via the AdS/CFT correspondence 
to the anomalous dimensions of a special class of
planar, large $R$-charge operators in ${\cal N}=4$ super Yang-Mills (SYM) 
theory \cite{Berenstein:2002jq,Witten:1998qj,Gubser:1998bc}.  This observation sparked a 
number of detailed tests of AdS/CFT which go beyond the supergravity limit 
of the string theory.  Many of these investigations have probed simplifying
limits of Maldacena's original proposal equating IIB superstring theory
on $AdS_5\times S^5$ with ${\cal N}=4$ SYM theory in four flat spacetime dimensions
\cite{Maldacena:1997re}.
It has been realized in the course of these studies that both the string
and gauge theory sides of this duality harbor integrable structures
associated with quantum spin chains.
The discovery of such structures is tantalizing, as it indicates 
progress toward finding exact solutions within certain sectors of the correspondence.
A more immediate benefit is that the arsenal of techniques associated with
the Bethe ansatz methodology has been useful for simplifying the computation
of both string energies and gauge theory operator dimensions.
Building on previous investigations (see, e.g., \cite{Berenstein:2002zw,Chen:2004mu,DeWolfe:2004zt,Lee:2002cu,Stefanski:2003qr,Susaki:2004tg,Chen:2004yf,Balasubramanian:2002sa,Takayanagi:2002je,Imamura:2002wz,Skenderis:2002wx,Skenderis:2002ps}), 
we study the integrability of open string states 
and explore the matchup with corresponding operator dimensions
in a dual defect conformal field theory.  This allows us to speculate on 
the role of a particular subset of non-perturbative interactions that have been 
conjectured to exist in the planar (large-$N_c$) limit of ${\cal N}=4$ SYM theory.

In this paper we will focus on a branch of work
involving the so-called near-pp-wave limit of the string theory
\cite{Parnachev:2002kk,Callan:2003xr,Callan:2004uv,Callan:2004ev,Swanson:2004mk,McLoughlin:2004dh,Swanson:2004qa}.   It was shown in 
\cite{Blau:2001ne,Blau:2002dy,Blau:2002mw} that the 
ten-dimensional pp-wave geometry is a consistent supergravity background
that can be realized as a large-radius Penrose 
limit of $AdS_5\times S^5$.  Finite-radius corrections to this limit
induce interactions on the string worldsheet that lift the highly degenerate 
energy spectrum on the pp-wave.  These near-pp-wave perturbations
were computed for the full supersymmetric string theory in 
\cite{Callan:2003xr,Callan:2004uv}, where the complete energy spectrum of 
two-impurity string states (characterized as having two mode excitations on the 
worldsheet) was compared in this limit to a corresponding set of operator dimensions 
computed in a perturbative regime of the gauge theory.
The string and gauge theory were found to exhibit a remarkably intricate agreement 
at one- and two-loop order in the 't~Hooft coupling $\lambda = g_{\rm YM}^2 N_c$, 
but suffer from a general mismatch at three-loop order and beyond.
This pattern of agreement at low orders in 
perturbation theory followed by higher-order disagreement
has been shown to repeat itself for the larger and more detailed
spectrum of three-impurity string states in \cite{Callan:2004ev}, and for the
complete set of $N$-impurity spectra in \cite{McLoughlin:2004dh}.
The sudden three-loop disagreement has also appeared in
related studies of the AdS/CFT correspondence involving
the comparison of semiclassical extended string configurations in anti-de~Sitter 
backgrounds with dual gauge theory operators 
\cite{Beisert:2003ea,Tseytlin:2003ii,Arutyunov:2004xy,Minahan:2004ds}.

\begin{figure}[htb]
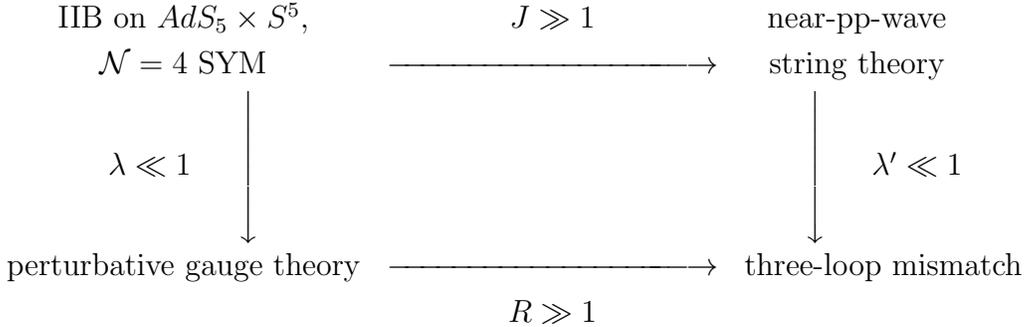

\begin{center}
\begin{eqnarray}
\begin{array}{ccc}
{\rm IIB\ on}\ AdS_5\times S^5,	 &  J\gg 1  & 	\kern-20pt \textrm{near-pp-wave}\\
{\cal N}=4\ {\rm SYM}	& \ext\ext\ext\longrightarrow &  \kern-20pt \textrm{string theory} \\
\lambda \ll 1 
\left. 
\phantom{{{{\int \over \int } \over \int}\over \int}\over \int}
\right\downarrow & & 
\left\downarrow
\phantom{{{{\int \over \int } \over \int}\over \int}\over \int}
\right. \lambda^\prime \ll 1   \\
\textrm{perturbative gauge theory}
	& \ext\ext\ext\longrightarrow &	 \textrm{three-loop mismatch}\\
	& R \gg 1 & 	
\end{array} \nonumber
\end{eqnarray}
\caption{The ordering of limits in the comparison of near-pp-wave string theory with 
	perturbative gauge theory}
\label{fig1}
\end{center}
\end{figure}

The matching procedure
between string and gauge theory in this setting is presented schematically 
in figure~\ref{fig1}.
The prevailing explanation for the higher-loop mismatch is that the comparison
of string energies with operator dimensions derived in the perturbative $\lambda$
expansion suffers from an order-of-limits problem 
\cite{Beisert:2004hm,Klebanov:2002mp,Serban:2004jf}.  
In the upper left-hand region we begin with the 
conjectured equivalence of IIB superstring theory on $AdS_5\times S^5$
and ${\cal N}=4$ SYM theory.  On the string theory side of the duality
one first takes a large-radius Penrose limit by boosting string states 
in $AdS_5\times S^5$ to large angular momentum $J$ 
along an equatorial geodesic in the $S^5$ subspace.  This large-$J$ limit
is combined with the 't~Hooft large-$N_c$ limit, which is taken such that 
the ratio $N_c/J^2$ remains fixed and finite.
For comparison with perturbative gauge theory, the large-$J$
Penrose limit is followed by an expansion in small 
$\lambda' = g_{\rm YM}^2 N_c / J^2$ (the so-called modified 't~Hooft coupling), 
leading to the lower right-hand region in 
figure~\ref{fig1}.  On the gauge theory side, operator dimensions are
first derived perturbatively in the small-$\lambda$ expansion.
The string angular momentum maps to the scalar component $R$ of the $SU(4)$
$R$-charge, so comparison with string theory 
requires a large-$R$ expansion, keeping $N_c/R^2$ fixed.  
This leads, via the left-hand route in figure~\ref{fig1},
to the matchup with string energies in the lower right-hand corner.
The outstanding question is whether the large-$J$ (or large-$R$) 
and small-$\lambda$ (or small-$\lambda'$) limits are commutative, and
whether non-commutativity can account for the mismatch at three-loop order
in $\lambda$.    

\begin{figure}[htbp]
  \begin{center}
    \mbox{
      \subfigure[One-loop order]{\scalebox{0.4}
{\includegraphics[width=5in,height=1.7in,angle=0]{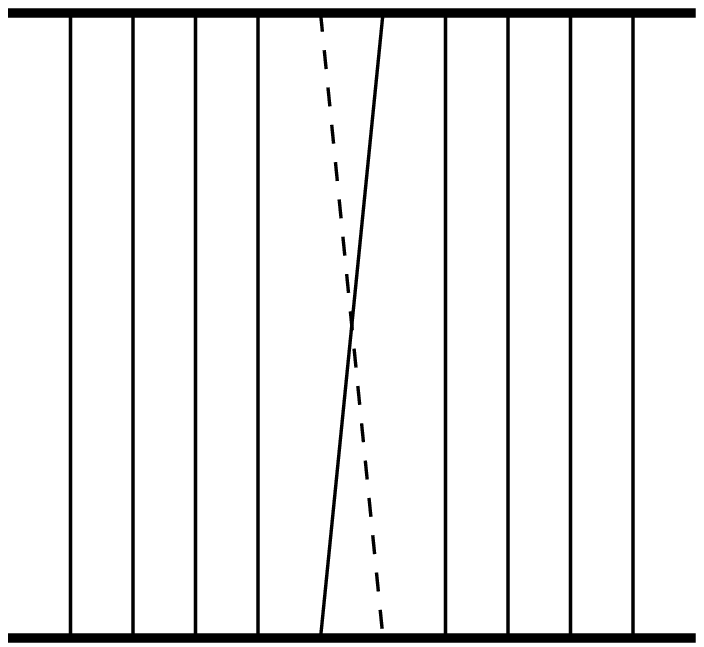}  }} \quad
      \subfigure[Two-loop order]{\scalebox{0.4}
{\includegraphics[width=5in,height=1.7in,angle=0]{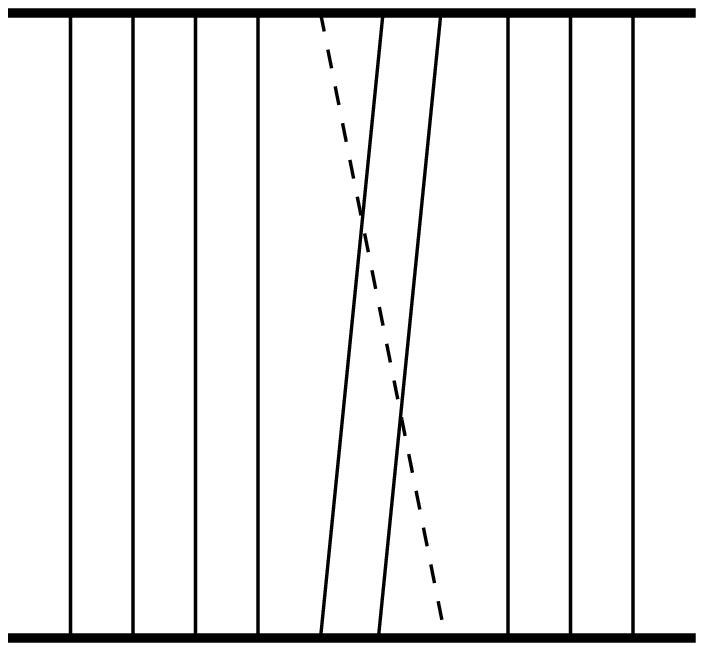}  }} \quad
      \subfigure[$n$-loop order]{\scalebox{0.4}
{\includegraphics[width=5in,height=1.7in,angle=0]{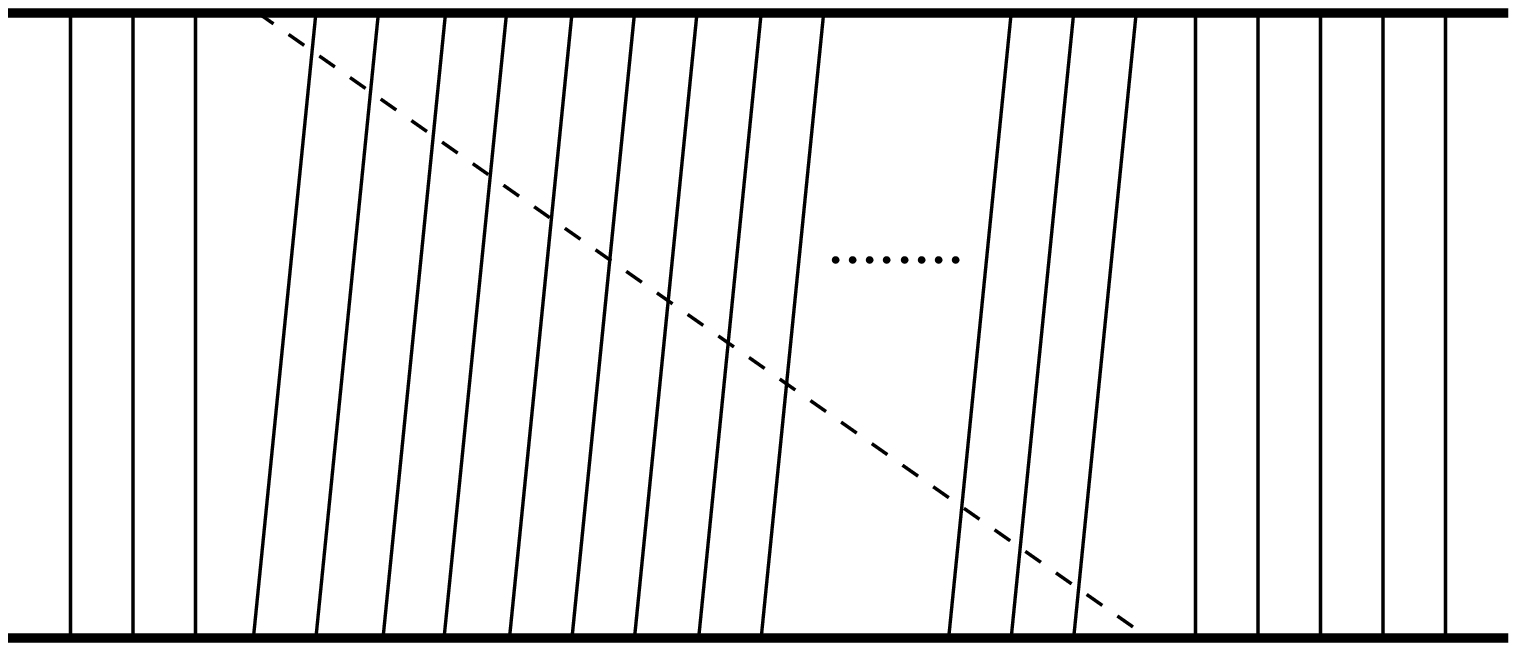}  }} 
      }
    \caption{Feynman diagrams at (a) one-, (b) two- and (c) $n$-loop order in $\lambda = g_{\rm YM}^2 N_c$}
    \label{nloop}
  \end{center}
\end{figure}

In the spin chain picture of the gauge theory \cite{Minahan:2002ve}, single-trace operators are
identified with cyclic lattice configurations of interacting spins. 
The excitation of magnon states on the spin lattice corresponds to 
the insertion of $R$-charge impurities in the trace; this defines a basis
on which the dilatation operator, or spin-chain Hamiltonian, can be diagonalized.  
At one-loop order in $\lambda$ the dilatation operator can be shown to 
mix only neighboring spins on the lattice
(i.e.,~dilatations of single-trace operators can only mix fields that are 
adjacent in the trace), and the corresponding statement at $n^{\rm th}$ 
order is that the $n$-loop Hamiltonian can mix fields on the lattice that 
are separated by at most $n$ lattice sites (see figure~\ref{nloop}).  
Within certain sectors of ${\cal N}=4$ SYM theory,
one strategy for extracting operator dimensions at higher loop order in $\lambda$
has been to derive the appropriate $n$-loop spin chain Hamiltonian 
by enumerating all possible interaction terms 
of maximum length $n$ and fix the coefficients of these terms using symmetry
constraints (or otherwise) handed down from the gauge theory 
\cite{Beisert:2003tq,Beisert:2003ys,Beisert:2004ry,Beisert:2004yq}.  
One way in which the limits in figure~\ref{fig1} could be non-commutative 
in this context is that this procedure may
neglect the contribution of interactions on the spin lattice that have 
a range greater than the total length of the lattice itself \cite{Beisert:2004hm}.  
The effect of these interactions would be apparent in any calculation that
attempts to apply asymptotic Bethe equations (derived for long chains)
to spin chains of finite length.
This is precisely the situation encountered when comparing finite $R$-charge corrections 
near the BMN limit to near-pp-wave corrections to the string energy spectrum.
The suggestion is that if one were able to eliminate such ultra long-range effects, 
complete agreement with corresponding string predictions would be obtained.

We attempt to shed light on the specific contribution 
of wrapping\footnote{To be certain, we use the term `wrapping' here to 
specify interactions which literally wrap around the lattice.}  
interactions by studying the matchup of open string states with 
corresponding gauge theory operators.  We formulate a prescription
for determining long-range, open-chain Bethe ans\"atze based on the scattering
matrices that appear in the analogous closed-chain equations.
On the string side we test our formulas by computing the full, 
all-loop energy spectrum of open string states in the near-pp-wave limit 
and comparing with results obtained by direct
quantization and diagonalization of the string lightcone Hamiltonian.
The string states of interest arise from the IIB theory on $AdS_5\times S^5$
with a $D5$ brane wrapping an $AdS_4\times S^2$ subspace \cite{DeWolfe:2001pq,Lee:2002cu}.  
The spacetime
dimensions occupied by both the $D3$ and $D5$ branes are indicated in 
table~\ref{dims}.  
\begin{table}[ht!]
\begin{eqnarray}
\begin{array}{|c|cccccccccc|}
\hline
 & 0 & 1& 2& 3& 4& 5& 6& 7& 8& 9 \\
\hline
D3& \times & \times& \times& & & & & & & \times  \\
D5& \times& \times& \times& \times& \times& \times& & & &  \\
\hline
\end{array} \nonumber
\end{eqnarray}
\caption{Extended dimensions of $D3$ and $D5$ branes in the ten-dimensional background}
\label{dims}
\end{table}
Alternatively, one may study an $AdS_5\times S^5/Z_2$ space,
which is the near-horizon geometry of a large number of $D3$-branes 
at an orientifold 7-plane in type ${\rm I}'$ theory \cite{Berenstein:2002zw}.
(The CFT dual is an $Sp(N)$ gauge theory.)  For the purposes of this paper, however, 
we find it sufficient to focus on the $D3$-$D5$ system, whose dual description
is an ${\cal N}=2$ defect conformal field theory.
We apply our Bethe ansatz on the CFT side of the duality by using the closed 
(or protected\footnote{To avoid confusion with open and closed strings, we will
refer to the closed, non-mixing sectors of the string and gauge theory as protected
sectors.}) $\su(2)$ sector of ${\cal N}=4$ SYM theory to derive 
an open-chain, long-range Bethe ansatz for an $\su(2)$ sector of operators 
in the ${\cal N}=2$ defect theory.

Let us note here that the mapping of gauge theory physics to the dynamics of integrable open spin 
chains has been studied in various different settings in the literature.
The authors of \cite{Chen:2004mu}, for example, studied the 
$Sp(N)$ superconformal theory containing one
hypermultiplet in the antisymmetric representation and four in the 
fundamental.  This theory had previously been studied in the BMN limit in
\cite{Berenstein:2002zw}, and its mixing matrix was shown in \cite{Chen:2004mu}
to map to the Hamiltonian of an integrable open spin chain.   
Similarly, DeWolfe and Mann \cite{DeWolfe:2004zt} considered a general 
defect conformal field theory \cite{DeWolfe:2001pq} 
(derived as a perturbation to ${\cal N}=4$
SYM theory) and demonstrated that integrable structures again emerge within
certain sectors.  
This system, which we focus on in this paper, was also examined in
\cite{Lee:2002cu},  where comparisons were carried out at one-loop 
order between the full pp-wave limit of the string theory and the full 
BMN limit of the ${\cal N}=2$ gauge theory (see also reference \cite{DeWolfe:2001pq}).  
We also note that 
analyses of semiclassical open string states and corresponding 
gauge theory operator dimensions were presented in 
\cite{Stefanski:2003qr,Susaki:2004tg,Chen:2004yf}.  The reader is referred to
\cite{Balasubramanian:2002sa,Takayanagi:2002je,Imamura:2002wz,Skenderis:2002wx,Skenderis:2002ps,Naculich:2002fh} for further related investigations.

In section~\ref{strings} we establish notation, review the derivation of the bosonic 
interacting string Hamiltonian in the near-pp-wave limit and compute energy 
spectra of open string states.  This is carried out for a 
protected $\su(2)$ sector of bosonic symmetric-traceless 
open strings in the $S^5$ subspace with Dirichlet boundary conditions, 
and for an analogous $\Sl(2)$ sector in the $AdS_5$ subspace with Neumann boundary conditions.
For pedagogical reasons, we treat separately the distinct cases of 
states with completely unequal mode
excitations and those with mode numbers that are allowed to overlap (although we carry this
out only for the $\su(2)$ sector).  
In section~\ref{StringBE} we formulate a long-range quantum Bethe ansatz for
open strings based on long-range scattering matrices developed for
corresponding sectors of closed string states in the pure $AdS_5\times S^5$
background.
We solve our resulting Bethe equations in the near-pp-wave limit for general 
open string states in both the Dirichlet $\su(2)$ and Neumann $\Sl(2)$ sectors,
and we compare our results with corresponding energy spectra computed
directly from the string theory in section~\ref{strings}.
In section~\ref{SYMBA} we apply our long-range Bethe ansatz to the $\su(2)$
and $\Sl(2)$ sectors of the dual ${\cal N}=2$ defect conformal field theory.  The
operators of interest are not gauge-traced, and the
open-chain Bethe ans\"atze should be free of any interference from 
wrapping interactions.  Upon solving the Bethe equations in the near-BMN
limit, we obtain agreement between operator dimensions in these sectors 
and the corresponding $\su(2)$ and $\Sl(2)$ string energy predictions at one and two-loop order 
in $\lambda$.  Once again, however, this agreement breaks down for both sectors 
at three-loop order and beyond.  We conclude in the final section with a summary and 
discussion of these results.

\section{Open string energy spectra near the pp-wave limit}
\label{strings}
The metric of $AdS_5\times S^5$ can be written in global coordinates as
\be
ds^2 = \widehat R^2 ( - {\rm cosh}^2 \rho~ dt^2 + d \rho^2 + {\rm sinh}^2
\rho~ d \Omega_3^2 + {\rm cos}^2  \theta~ d \phi^2 +  d \theta^2 +
{\rm sin}^2 \theta~ d \widetilde\Omega_3^2)~,
\label{adsmetric0}
\ee
where the common scale radius of both the $AdS_5$ and $S^5$ subspaces
is denoted by $\Rhat$, $t$ is the global time direction, and $d \Omega_3^2$
and $d \widetilde\Omega_3^2$ indicate separate three-spheres.
As in previous studies, we invoke the following
reparameterizations
\begin{eqnarray}
    \cosh\rho  =  \frac{1+z^2/4}{1-z^2/4}~, \qquad
    \cos\theta  =  \frac{1-y^2/4}{1+ y^2/4}\ ,
\end{eqnarray}
and work with the metric
\be
\label{metric0}
ds^2  = \widehat R^2
\biggl[ -\left({1+ \frac{1}{4}z^2\over 1-\frac{1}{4}z^2}\right)^2dt^2
        +\left({1-\frac{1}{4}y^2\over 1+\frac{1}{4}y^2}\right)^2d\phi^2
    + \frac{dz_k dz_k}{(1-\frac{1}{4}z^2)^{2}}
    + \frac{dy_{k'} dy_{k'}}{(1+\frac{1}{4}y^2)^{2}} \biggr]\ .
\ee
This version of the spacetime metric is useful when working with fermions
(see \cite{Callan:2004uv} for details), though
we will restrict ourselves to the bosonic sector of the string theory in the
present study.  The Penrose limit is reached by boosting string states to
lightlike momentum $J$ along an equatorial geodesic in the $S^5$ subspace.
Under the rescaling prescriptions
\be
\label{rescale}
    t \rightarrow x^+~,
\qquad
    \phi \rightarrow x^+ + \frac{x^-}{\widehat R^2}~,
\qquad
    z_k \rightarrow \frac{z_k}{\widehat R}~,
\qquad
    y_{k'} \rightarrow \frac{y_{k'}}{\widehat R}~,
\ee
this limit is reached by taking $\widehat R\to\infty$.
The $S^5$ angular momentum $J$ is thereby related to the scale radius $\Rhat$
by 
\be
p_-\Rhat^2 = J\ ,
\ee
and the lightcone momenta are given by 
\be
-p_+ = \Delta-J~, \qquad -p_-=i\partial_{x^-} = \frac{i}{\Rhat^2}\partial_\phi 
	= -\frac{J}{\Rhat^2}\ .
\ee
The transverse Cartesian coordinates $z_k$ and $y_{k'}$ span
an $SO(4)\times SO(4)$ subspace, with $z_k$ lying in $AdS_5$ 
($k \in 1,\ldots ,4$) and $y_{k'}$ in the $S^5$ subspace
($k' \in 5,\ldots ,8$).
In the Penrose limit $p_-$ is held fixed as $J$ and $\Rhat$ become infinite. 
We also work in the planar limit, where the number of colors (or number of $D3$ branes) 
$N_c$ becomes large, but the quantity $N_c/J^2$ is held fixed.
The lightcone momentum $p_-$ is equated via the AdS/CFT dictionary with
\be
p_- = \frac{1}{\sqrt{\lambda'}} = \frac{J}{\sqrt{g_{\rm YM}^2 N_c}}\ ,
\ee
where, as noted in the introduction,  
$\lambda'$ is known as the modified 't~Hooft coupling, 
and $J$ is equated on the CFT 
side with the scalar $R$-charge $R$.

For reasons described in \cite{Callan:2003xr,Callan:2004uv}, 
the lightcone coordinates in eqn.~(\ref{rescale}) admit
many simplifications, the most important of which is
the elimination of normal-ordering contributions to the 
lightcone Hamiltonian (at least to the order of interest).  
In terms of these coordinates, 
we obtain the following large-$\Rhat$ expansion of the metric:
\be
\label{expndmet0}
ds^2 & = &
	2\,{dx^+}{dx^-} + {dz}^2 + {dy }^2  -
        \left( {z }^2 + {y }^2 \right) ({dx}^+)^2 
\nn\\
&&\kern-20pt
	+ \frac{1}{\widehat R^2} \left[- 2 y^2 dx^- dx^+
	+\frac{1}{2} \left( {y }^4 - {z }^4 \right) (dx^+)^2 + \left(d x^-\right)^2
    	+ \frac{1}{2}z^2 dz^2 - \frac{1}{2} y^2 dy^2 \right]
   	+ O(1/\widehat R^2)\ ,
\nn\\
&&
\ee
where the coordinates $x^A$ span an $SO(8)$ subspace, with $A\in 1,\ldots,8$.
The pp-wave metric emerges at leading order, and the string theory on that 
background is free \cite{Metsaev:2001bj,Metsaev:2002re}.  The background curvature 
correction appearing at $O(1/\Rhat^2)$ induces a set of interaction
corrections to the free string spectrum in the Penrose limit.  As noted in the introduction,
these corrections lift the degeneracy of the free theory on the pp-wave, 
allowing for much more detailed comparisons of string energy spectra 
with gauge theory anomalous dimensions.

In the Penrose limit, the lightcone Green-Schwarz action for open superstrings 
takes the form 
\be
\label{ppwavact}
S_{\rm pp} = \frac{1}{2\pi} \int d\tau
\int_0^{2\pi } d \sigma ({\cal L}_B+{\cal L}_F)\ ,
\ee
where
\begin{eqnarray}
\label{LB}
{\cal L}_B &=& \frac{1}{2} \left[ ( \dot x^A)^2
-  (x^{\prime A})^2 - (x^A)^2\right]~,
\\
\label{LF}
{\cal L}_F &=& i \psi^{\dagger} \dot\psi + \psi^{\dagger} \Pi
\psi +\frac{i}{2} ( \psi \psi^{\prime} + \psi^{\dagger}
\psi^{\prime\dagger} )~.
\end{eqnarray}
(We have set $\alpha'=1$.)
The fields $\psi_\alpha$ are eight-component complex spinors formed from
two Majorana-Weyl $SO(9,1)$ spinors of equal chirality, and $\Pi$ is 
defined in terms of eight-dimensional $SO(8)$ Dirac gamma matrices by
$\Pi \equiv \gamma^1\bar \gamma^2 \gamma^3 \bar \gamma^4$ (see \cite{Callan:2004uv} for
further details).  The shorthand notation $\dot x^A$ and 
${x'}^A$ denotes the worldsheet derivatives $\del_\tau x^A$ and $\del_\sigma x^A$,
respectively.  The pp-wave lightcone Hamiltonian is easily derived from
eqns.~(\ref{LB}) and (\ref{LF}):
\be
H_{\rm pp} = \frac{p_-}{2}(x^A)^2 
	+ \frac{1}{2p_-}\left[(p_A)^2 + ({x'}^A)^2 \right]
	+ i\rho \Pi \psi + \frac{i}{2}\psi\psi' - \frac{i}{2p_-^2}\rho\rho'\ ,
\ee 
where the fields $p_A$ and $\rho_\alpha$ are conjugate to $x^A$ and $\psi_\alpha$,
respectively.

Including the first finite-radius curvature correction to the spacetime metric
in eqn.~(\ref{expndmet0}) gives rise to an interacting Hamiltonian appearing at
$O(1/\Rhat^2)$ in the large-radius expansion.  This Hamiltonian was computed and analyzed 
extensively in \cite{Callan:2003xr,Callan:2004uv,Callan:2004ev,McLoughlin:2004dh}.  
The bosonic sector, labeled by $H_{\rm BB}$, is 
quartic in fields and mixes purely bosonic string states:
\be
\label{Hpurbos}
{H}_{\rm BB} & = & \frac{1}{\Rhat^2}\biggl\{
	\frac{1}{4p_-}\left[ -y^2\left( p_z^2 + {z'}^2 + 2{y'}^2\right)
	+ z^2\left( p_{y}^2 + {y'}^2 + 2{z'}^2 \right)\right]
	+ \frac{p_-}{8}\left[ (x^A)^2 \right]^2
\nn\\
& & 	- \frac{1}{8p_-^3}\left[  \left[ (p_A)^2\right]^2 + 2(p_A)^2({x'}^A)^2 
	+ \left[ ({x'}^A)^2\right]^2 \right]
	 + \frac{1}{2p_-^3}\left({x'}^A p_A\right)^2
	\biggr\}\ .
\ee
We have used the shorthand symbols $p_z$ and $p_y$ to denote bosonic momentum 
fields in the $AdS_5$ and $S^5$ subspaces, respectively.    
Since we will not be dealing with fermions, the string states of interest 
are formed by acting with bosonic creation operators $a_n^{A\dag}$ on a vacuum state 
$\ket{J}$ carrying $J$ units of angular momentum in the $S^5$ subspace: 
\be
\Ket{N;J} \equiv	\underset{N}{\underbrace{a_{n_1}^{A_1\dag}a_{n_2}^{A_2\dag} 
	\cdots a_{n_{N}}^{A_{N}\dag}}} \ket{J}\ .
\nn
\ee
Such states are generically referred to as $N$-impurity bosonic string states.

In \cite{Callan:2003xr,Callan:2004uv,Callan:2004ev} it was demonstrated that 
there are certain special sectors of the full interacting Hamiltonian that 
completely decouple from the theory.  Among these are two bosonic sectors consisting 
of pure-boson states restricted to either the $AdS_5$ or $S^5$ 
subspaces and projected onto symmetric-traceless irreps of spacetime $SO(4)$, 
plus one sector comprised of purely fermionic excitations symmetrized
in spinor indices (symmetrized spinors survive in this setting because
they come with additional mode-number labels).
For low impurity number, these sectors
are conveniently labeled according to their transformation properties under the residual
$SO(4)_{AdS}\times SO(4)_{S^5}$ symmetry that survives in the Penrose (or near-Penrose) limit.
Employing the $SU(2)^2\times SU(2)^2$ notation used extensively in 
\cite{Callan:2003xr,Callan:2004uv,Callan:2004ev}, the two-impurity bosonic sectors 
transform as $({\bf 3,3;1,1})$ and $({\bf 1,1;3,3})$, while the two-impurity 
fermionic sectors are labeled
by $({\bf 3,1;3,1})$ and $({\bf 1,3;1,3})$.  (The three-impurity version of this
irrep decomposition is given in \cite{Callan:2004ev}.)  On the gauge theory side the 
symmetric-traceless $AdS_5$ $({\bf 3,3;1,1})$ bosons map to a closed sector of
the dilatation generator invariant under an $\Sl(2)$ subalgebra of the full superconformal
algebra $\psu(2,2|4)$.  The protected sector of $({\bf 1,1;3,3})$ bosonic $S^5$ string
states maps to an $\su(2)$ sector, and the $({\bf 3,1;3,1}) + ({\bf 1,3;1,3})$ fermions
correspond to an $\su(1|1)$ sector.  We will use these algebraic labels when we
refer to these decoupled subsectors in both the string theory and the gauge theory.

Since we are interested in general multi-impurity states within these protected 
subsectors, we will restrict ourselves in the $\Sl(2)$ sector to open strings
with Neumann boundary conditions, and the $\su(2)$ sector will be comprised of
open strings with Dirichlet boundary conditions.
The available bosonic worldsheet fields are indicated in table~\ref{ND},
where, for example, $z_N^j$ indicates fields excited in the $AdS_5$ subspace
with Neumann boundary conditions, and $y_D^{j'}$ are $S^5$ fluctuations with
Dirichlet boundary conditions.

\begin{table}[ht!]
\begin{eqnarray}
\begin{array}{|cc|cccc|cccc|}
\hline
+ & - & 1& 2& 3& 4& 5& 6& 7& 8  \\
\hline
x^+ & x^- & z_N^j & z_N^j & z_N^j & z_D^j & y_N^{j'} & y_D^{j'} & y_D^{j'} & y_D^{j'}   \\
\hline
\end{array} \nonumber
\end{eqnarray}
\caption{Neumann and Dirichlet directions on the $SO(4)_{AdS}\times SO(4)_{S^5}$ transverse subspace}
\label{ND}
\end{table}

\subsection{Dirichlet $SO(4)_{S^5}$ $(\su(2))$ sector}
Since the goal is to compute $O(1/J)$ 
interaction corrections to the free energy spectrum in the pp-wave limit,
we find it convenient to isolate such corrections according to the energy expansion
\be
E(\{n_j\},N,J) = \sum_{j=1}^N\sqrt{1+n_j^2\lambda'/4}
	+ \delta E(\{n_j\},N,J) + O(1/J^2)\ .
\ee
The near-pp-wave energy spectra we are interested in are thus collected into
the $O(1/J)$ correction $\delta E(\{n_j\},N,J)$, which generically depends
on the set of mode numbers $\{n_j\}$ carried by the corresponding (unperturbed)
energy eigenstate, the total number of worldsheet mode excitations $N$, and
the string $S^5$ angular momentum $J$.

We will focus first on a set of $N$-impurity bosonic open string states 
in the protected $\su(2)$ sector.  
We form symmetric-traceless (in $SO(4)_{S^5}$ indices) states by combining 
excitations in the Dirichlet $(y^6,y^7,y^8)$ $S^5$ directions:
\be
\Ket{N;J}_D = {{a_{n_1}^{(j'_1\dag}a_{n_2}^{j'_2\dag} 
	\cdots a_{n_{N}}^{j'_{N})\dag}}} \ket{J}\ ,
\label{s5states}
\ee
where the $S^5$ labels $j'$ indicate the $(y^6,y^7,y^8)$ 
directions and contributions to the traced state are understood to be absent.

The Fourier expansion of bosonic fluctuations in the Dirichlet directions
is given by \cite{Dabholkar:2002zc}
\be
x^A_D(\tau,\sigma) = \sum_{n=1}^\infty \frac{i}{\sqrt{\omega_n}}
		\sin\left(\frac{n\sigma}{2}\right)
		\left(  a_n^{A}e^{-i\omega_n \tau} - a_n^{A\dag}e^{i\omega_n \tau} \right)\ .
\ee
The equations of motion in the pp-wave limit
\be
\ddot x^A - {x''}^A + p_-^2 x^A = 0
\ee
are therefore satisfied by the dispersion relation
\be
\label{disp}
\omega_n = \sqrt{p_-^2 + (n/2)^2}\ ,
\ee
where the mode index $n$ is integer-valued and the oscillators $a_n^A$ and 
$a_n^{A\dag}$ satisfy the usual commutation 
relations: $\left[ a_n^A, a_m^{B\dag} \right] = \delta_{n,m}\delta^{A,B}$.
We are interested in computing diagonal matrix elements of $H_{\rm BB}$ 
between physical string states which, in first-order perturbation theory,
will involve equal numbers of creation and annihilation operators 
(i.e.,~we need not be concerned with sectors of $H_{\rm BB}$ that mix states
with different numbers of impurities).  We may therefore restrict to terms 
appearing in $H_{\rm BB}$ with two creation and two annihilation operators.

Following the approach in \cite{McLoughlin:2004dh}, we simplify the 
projection onto symmetric-traceless $S^5$ string states by forming the following 
bosonic oscillators:
\be
a_n  \equiv \frac{1}{\sqrt{2}}\left(a_n^6 + i a_n^7\right)~, \qquad 
\bar a_n \equiv \frac{1}{\sqrt{2}}\left(a_n^6 - i a_n^7\right)~. 
\label{oscdef}
\ee
Generic matrix elements of the form
\be
\bra{J} a_{n_1}a_{n_2}\cdots a_{n_{N}} ( H_{\rm BB} ) 
	a_{n_1}^\dag a_{n_2}^\dag \cdots a_{n_{N}}^\dag \ket{J}
\label{ME}
\ee 
therefore select out excitations in the $(y^6_D,y^7_D)$ plane and make the 
symmetric-traceless projection manifest.  Restricting to the Dirichlet 
directions, we are also free to project onto the $(y^6_D,y^8_D)$ and 
$(y^7_D,y^8_D)$ planes.  We exclude the $y^5_N$ direction because we
do not want to mix Neumann and Dirichlet boundary conditions (although,
in general, such states are certainly allowed), and states
in the $S^5$ subspace built strictly from $y^5_N$ excitations
would be projected out by the traceless condition.

As described in \cite{McLoughlin:2004dh}, the oscillator expansion of $H_{\rm BB}$
admits two generic structures: one characterized by a contraction of
$SO(4)$ indices attached to pairs of creation or annihilation operators
\be
a_{n}^{\dag A} a_{m}^{\dag A} a_{l}^{B}a_{p}^{B}\ , \nn
\ee
and one containing pairs of contracted creation and annihilation operators 
\be
a_{n}^{\dag A} a_{l}^{\dag B} a_{m}^{A}a_{p}^{B}\ .  \nn
\ee
(The indices $\{n,l,m,p\}$ are mode numbers appearing in the mode expansion
of $H_{\rm BB}$.)
Expanding in the fields defined in eqn.~(\ref{oscdef}), we obtain
\be
\label{osc1}
a_{n}^{\dag A} a_{m}^{\dag A} a_{l}^{B}a_{p}^{B}\Bigr|_{(6,7)} &=& 
	\bigl( a_{n}^\dag\,\bar a_{m}^\dag + \bar a_{n}^\dag\, a_{m}^\dag \bigr)
	\bigl( a_l\,\bar a_p + \bar a_l\, a_p\bigr)\ ,
\\
a_{n}^{\dag A} a_{l}^{\dag B} a_{m}^{A}a_{p}^{B}\Bigr|_{(6,7)} &=& 
	\bar a_{n}^\dag\, \bar a_{l}^\dag\, \bar a_m\, \bar a_p
	+ a_{n}^\dag\, a_{l}^\dag\, a_m\, a_p\ . 
\label{osc2}
\ee
Only the second term in eqn.~(\ref{osc2})
will contribute to the matrix elements in (\ref{ME}).  We may therefore 
simplify the calculation of the energy spectrum in this sector of the theory
by projecting the Hamiltonian onto terms containing only the oscillator
structure $a_{n}^\dag a_{l}^\dag a_m a_p$.

To begin we will compute energy eigenvalues for the simplest 
eigenstates in which all mode numbers of the symmetric-traceless state 
\be
a_{n_1}^\dag a_{n_2}^\dag \cdots a_{n_{N}}^\dag \ket{J} \nn
\ee
are taken to be unequal $(n_1 \neq n_2 \neq \cdots \neq n_N)$.
Between these states, the interaction Hamiltonian will admit 
matrix elements whose structure is defined by
\be
&&\bra{J}a_{n_1}a_{n_2}\ldots a_{N_B} ( a_{n}^\dag a_{l}^\dag a_m a_p ) 
	a_{n_1}^\dag a_{n_2}^\dag \ldots a_{N_B}^\dag \ket{J}
\nn\\
&&\kern+60pt
	= \frac{1}{2}
	\sum_{j,k=1 \atop j\neq k}^{N} 
	\Bigl(
	\delta_{n_j,n}\,\delta_{n_k,l}\,\delta_{n_j,m}\,\delta_{n_k,p}
	+\delta_{n_j,n}\,\delta_{n_k,l}\,\delta_{n_k,m}\,\delta_{n_j,p}
\nn\\
&&\kern+110pt
	+\delta_{n_j,l}\,\delta_{n_k,n}\,\delta_{n_j,m}\,\delta_{n_k,p}
	+\delta_{n_j,l}\,\delta_{n_k,n}\,\delta_{n_k,m}\,\delta_{n_j,p}
	\Bigr)\ .
\label{delME1}
\ee
The appropriate energy eigenvalue in this sector can thus be computed
by attaching coefficients of $a_{n}^\dag a_{l}^\dag a_m a_p$
in $H_{\rm BB}$ to the matrix element structure in eqn.~(\ref{delME1}).
The $O(1/J)$ energy correction for this protected $\su(2)$ sector of symmetric-traceless
$S^5$ open string states is thereby found, for completely unequal mode indices, to be:
\be
\label{nonconfluentE}
\delta E_{{S^5}}(\{n_i\},N,J) &=& -\frac{1}{8J} \sum_{j,k =1 \atop j\neq k}^N
	\frac{n_k^2 + n_j^2 \lambda' \omega_{n_k}^2}{\omega_{n_j}\omega_{n_k}}\ .
\ee
By expanding in small $\lambda'$
\be
\delta E_{{S^5}}(\{n_i\},N,J)
	&=& \frac{1}{J}\sum_{j,k=1\atop j\neq k}^N 
	\biggl\{
	-\frac{1}{8}(n_j^2+n_k^2)\lambda'
	+\frac{1}{64}(n_j^4+n_k^4)(\lambda')^2
\nn\\
&&	-\frac{1}{1024}(3n_j^6 + n_j^4 n_k^2 + n_j^2 n_k^4 + 3 n_k^6 )(\lambda')^3
	+ O((\lambda')^4) \biggr\}\ ,
\ee
it is easy to see that the spectrum for these states exhibits the
expected property that at $m^{\rm th}$ order in the expansion the 
energy scales with $2m$ factors of the mode numbers $\{n_i\}$.

We can extend this result to the most general set of states in this sector, 
formed by a set of $a_n^\dag$ creation oscillators grouped into $M$ subsets, with all
mode indices equal within these subsets:
\be
\frac{\left( a_{n_1}^\dag\right)^{N_{1}}}{\sqrt{N_{1}!}}
\frac{\left( a_{n_2}^\dag\right)^{N_{2}}}{\sqrt{N_{2}!}}
\cdots
\frac{\left( a_{n_M}^\dag\right)^{N_{M}}}{\sqrt{N_{M}!}}\ket{J}\ .
\nn
\ee
The $j^{th}$ subset contains $N_{j}$ oscillators with mode index $n_j$,
and the total impurity number is $N$, such that
\be
\sum_{j=1}^M N_{j} = N\ .
\ee
The matrix element of the oscillator structure $a_{n}^\dag\, a_{l}^\dag\, a_m\, a_p$ 
between these states is
\be
&&\kern-30pt
\bra{J} 
\frac{\left( a_{n_1} \right)^{N_{n_1}}}{\sqrt{N_{n_1}!}}
\cdots
\frac{\left( a_{n_M} \right)^{N_{n_M}}}{\sqrt{N_{n_M}!}}
\left( a_{n}^\dag\, a_{l}^\dag\, a_m\, a_p \right)
 \frac{\left( a_{n_1}^\dag\right)^{N_{n_1}}}{\sqrt{N_{n_1}!}}
\cdots
\frac{\left( a_{n_M}^\dag\right)^{N_{n_M}}}{\sqrt{N_{n_M}!}}
\ket{J}
\nn\\
&&\kern+00pt
	= \sum_{j=1}^M N_{n_j}(N_{n_j}-1)\,\delta_{n,n_j}\,\delta_{l,n_j}\,
	\delta_{m,n_j}\,\delta_{p,n_j}
	+\frac{1}{2}\sum_{j,k=1\atop j\neq k}^M N_{n_j} N_{n_k}
	\Bigl(
	\delta_{n,n_k}\,\delta_{l,n_j}\,\delta_{m,n_k}\,\delta_{p,n_j}
\nn\\
&&\kern+40pt
	+\delta_{n,n_j}\,\delta_{l,n_k}\,\delta_{m,n_k}\,\delta_{p,n_j}
	+\delta_{n,n_k}\,\delta_{l,n_j}\,\delta_{m,n_j}\,\delta_{p,n_k}
	+\delta_{n,n_j}\,\delta_{l,n_k}\,\delta_{m,n_j}\,\delta_{p,n_k}
	\Bigr)\ .
\label{genME}
\ee
The energy shift at $O(1/J)$ for completely general $N$-impurity 
open string states in the Dirichlet $\su(2)$ sector of the theory
is therefore given by
\be
\label{confluentE}
\delta E_{S^5}(\{n_i\},\{N_{i}\},M,J)  &=&  -\frac{1}{8 J}\biggl\{
	\sum_{j=1}^M N_{j}(N_{j}-1) \frac{n_j^2(6+n_j^2\lambda')}{4 \omega_{n_j}^2}
\nn\\
&&\kern+40pt
	+ \sum_{j,k =1 \atop j\neq k}^M N_{j} N_{k}
	\frac{n_k^2 + n_j^2 \lambda' \omega_{n_k}^2}{\omega_{n_j}\omega_{n_k}}
	\biggr\}\ .
\ee
We see that the structure obtained for states with completely inequivalent
mode indices in eqn.~(\ref{nonconfluentE}) appears in the second term above.
This term can be thought of as the contribution to the energy from scattering
among excitations with differing mode numbers; the first term represents
scattering within subsets of equal mode numbers.
For eventual comparison with corresponding quantities in the dual gauge theory,
we expand eqn.~(\ref{confluentE}) in small $\lambda'$:
\be
\delta E_{S^5}(\{n_i\},\{N_{n_i}\},M,J) &=&
	 \frac{1}{J}\sum_{k=1}^M N_k(N_k-1) 
	\Bigl[ 
	-\frac{3}{16} n_k^2\lambda'
	+ \frac{1}{64}n_k^4(\lambda')^2-\frac{1}{256}n_k^6(\lambda')^3 \Bigr]
\nn\\
&&
\kern-20pt
	+ \frac{1}{J}\sum_{k,j=1\atop j\neq k}^M N_j N_k 
	\Bigl[
	-\frac{1}{8}(n_j^2+n_k^2)\lambda' + \frac{1}{64}(n_j^4+n_k^4)(\lambda')^2
\nn\\
&&
\kern-20pt
	-\frac{1}{1024}(3n_j^6 + n_j^4 n_k^2 + n_j^2 n_k^4 + 3n_k^6)(\lambda')^3 
	 \Bigr] + O((\lambda')^4)\ .
\label{stringEXP}
\ee
Within each of the $M$ subsectors of overlapping mode numbers (labeled by the indices $j$ and $k$)
we again observe the desired result that the contribution to the energy
scales at $m^{\rm th}$ order in $\lambda'$ in proportion to $n_k^{2m}$.

\subsection{Neumann $SO(4)_{AdS}$ ($\Sl(2)$) sector}
From table~\ref{ND} we see that we can repeat the above calculation for
the protected $\Sl(2)$ sector of symmetric-traceless bosonic open string states excited
in the $AdS_5$ subspace.  Analogous to eqn.~(\ref{oscdef}) above, 
we can project onto this sector by forming the oscillators
\be
a_n  \equiv \frac{1}{\sqrt{2}}\left(a_n^1 + i a_n^2\right)~, \qquad 
\bar a_n \equiv \frac{1}{\sqrt{2}}\left(a_n^1 - i a_n^2\right)~,
\label{oscdef2}
\ee
and restricting to unperturbed string states of the form
\be
a_{n_1}^\dag a_{n_2}^\dag \cdots a_{n_{N}}^\dag \ket{J}\ . \nn
\ee
Here we are free to project onto the $(z^1_N,z^2_N)$, $(z^2_N,z^3_N)$ or
$(z^1_N,z^3_N)$ planes, and we choose to exclude $z^4_D$ to avoid mixing boundary 
conditions.
The states of interest carry Neumann boundary conditions,
and the the mode expansion is given by
\be
x^A_N(\tau,\sigma) = \sum_{n=0}^\infty \frac{i}{\sqrt{\omega_n}}
		\cos\left(\frac{n\sigma}{2}\right)
		\left(  a_n^{A}e^{-i\omega_n \tau} - a_n^{A\dag}e^{i\omega_n \tau} \right)\ .
\label{NBC}
\ee
(The equations of motion again require that $\omega_n$ satisfy the dispersion 
relation given in eqn.~(\ref{disp}).)

The computation of the energy shift at $O(1/J)$ for the $\Sl(2)$ sector of 
open string states follows in complete analogy with the $\su(2)$ sector described above.
We form general matrix elements of the form given in eqn.~(\ref{genME}) and 
attach coefficients of the Hamiltonian proportional to the oscillator 
structure $a_{n}^\dag\, a_{l}^\dag\, a_m\, a_p$, the only difference
being that this sector of the interacting Hamiltonian is defined in terms 
of the worldsheet excitations in eqn.~(\ref{NBC}).
Leaving out the computational details, 
we obtain the following energy shift for completely general open string 
states in the $\Sl(2)$ sector of the theory:
\be
\delta E_{AdS}(\{n_i\},\{N_i\},M,J)  &=&  
	-\frac{1}{32\,J}\biggl\{ 
	\sum_{k=1}^M N_k (N_k-1) \frac{n_k^2 (2+n_k^2\lambda')}{\omega_{n_k}^2}	
	+\sum_{j,k=1 \atop j\neq k}^M N_j N_k 
	\frac{n_j^2 n_k^2 \lambda'}{\omega_{n_j}\omega_{n_k}}
	\biggr\}\ .
\nn\\
&&
\label{EAdS}
\ee
For comparison with the gauge theory, we expand in $\lambda'$:
\be
\delta E_{AdS}(\{n_i\},\{N_i\},M,J)
	&=& 
	\frac{1}{J}\sum_{k=1}^M N_k(1-N_k) \Bigl[
	\frac{1}{16}n_k^2 \lambda'
	+\frac{1}{64}n_k^4 (\lambda')^2
	-  \frac{1}{256}n_k^6 (\lambda')^3 
	\Bigr]
\nn\\
&&\kern-00pt
	+\frac{1}{J}\sum_{j,k=1\atop j\neq k}^M N_j N_k 
	\Bigl[ -\frac{1}{32} n_j^2 n_k^2 (\lambda')^2 
	+ \frac{1}{256}n_j^2 n_k^2 (n_j^2 + n_k^2)(\lambda')^3
	\Bigr]
\nn\\
&&	+ O((\lambda')^4)\ .
\label{SL2EXP}
\ee
The result again breaks up into contributions from interactions among 
worldsheet excitations with both equal and inequivalent mode numbers.
There are also no contributions from zero-mode
fluctuations:  this is an important consistency check, since such corrections
are prohibited in general \cite{Callan:2003xr,Callan:2004uv}.
One may also expand in small $\lambda'$ to see that 
the scaling with mode numbers follows the same pattern 
demonstrated in eqns.~(\ref{nonconfluentE}) and (\ref{confluentE}) above.

\section{Open string Bethe equations}
\label{StringBE}
There has recently been a great deal of work exploring the existence and 
role of integrable structures in both type IIB string theory on $AdS_5\times S^5$
and ${\cal N}=4$ SYM theory. 
It was originally noticed, first for the bosonic theory \cite{Mandal:2002fs} 
and later for the full Green-Schwarz action 
\cite{Bena:2003wd}, that the 
classical coset sigma model of the string on $AdS_5\times S^5$ 
possesses an infinite set of mutually commuting conserved charges.  
In \cite{Arutyunov:2003rg}, Arutyunov and Staudacher used a set of 
B\"acklund transformations to construct a generating
function for these charges and matched their results to a corresponding 
generating function computed in the dual gauge theory.  
In \cite{Kazakov:2004qf}, the Riemann-Hilbert problem for 
the classical finite-gap solutions of the
sigma model, restricted to an $S^2\subset S^5$ subspace, was shown
to be equivalent to the classical limit of the corresponding gauge theory 
Bethe equations.  This result was extended to include the full sigma model in 
\cite{Kazakov:2004nh,Beisert:2004ag,Schafer-Nameki:2004ik,Beisert:2005bm}. 
This was accompanied by \cite{Berkovits:2004xu}, in which Berkovits used a
pure-spinor formalism to argue that that the tower of commuting conserved charges 
persists in the quantum theory (see also \cite{Swanson:2004qa} 
for an exploration of these charges in the quantum theory near
the BMN limit).  In \cite{Arutyunov:2004yx}, 
Arutyunov and Frolov were able to construct a Lax representation of the 
classical bosonic string Hamiltonian in a specific gauge; 
this result was extended by Alday, Arutyunov and Tseytlin
to the gauge-fixed physical superstring in \cite{Alday:2005gi}.
(The approach to studying higher conserved charges using monodromy matrices 
was related to earlier investigations involving B\"acklund 
transformations in \cite{Arutyunov:2005nk}.)
Finally, we note that studies of integrability for semiclassical rotating 
strings were extended beyond the planar limit in \cite{Peeters:2004pt,Peeters:2005pb}.

On the gauge theory side, the presence of integrable 
structures was originally pointed out
by Minahan and Zarembo \cite{Minahan:2002ve}, who showed that the action of the SYM
dilatation operator on single-trace operators in a protected $SO(6)$-invariant 
sector of the theory can be mapped to that of an integrable Hamiltonian 
acting on a closed Heisenberg spin chain.  
The eigenvalues of the Hamiltonian, which are identified with corresponding
operator dimensions, may thereby be obtained by solving a system of Bethe equations.
Since we aim to formulate and solve a set of Bethe equations for the open string states
examined in section~\ref{strings}, we will briefly review how this procedure 
is applied to the $\su(2)$ sector of ${\cal N}=4$ SYM theory at one-loop order in $\lambda$.
(We will return to the higher-loop treatment of the gauge theory 
in section~\ref{SYMBA} below.)

In the planar limit, single-trace operators in the $\su(2)$ sector 
are built from complex scalars, 
typically denoted by $Z$ and $\phi$.  The $Z$ fields
are charged under the $U(1)_R$ component of the $SO(6) \cong U(1)_R\times SO(4)$
decomposition of the full $SU(4)$ $R$-symmetry group, while the $\phi$ fields
carry zero $R$-charge and act as impurity insertions.
We obtain a basis of length-$L$ operators by forming field monomials 
that carry $N$ impurities and total $R$-charge equal to $R=L-N$:
\be
\tr(\phi^N Z^{L-N})\ , \qquad 
\tr(\phi^{N-1}Z\phi Z^{L-N-1})\ , \qquad
\tr(\phi^{N-2}Z\phi^2 Z^{L-N-1})\ , \qquad \ldots
\label{basis}
\ee
Unlike the $\so(6)$ sector originally studied by Minahan and Zarembo, 
which is protected from mixing at one-loop order in $\lambda$, the $\su(2)$ sector is 
protected at all orders in perturbation theory.\footnote{In other words, the
operators in eqn.~(\ref{basis}) will not mix with other operators
in the theory, and the anomalous dimension matrix can be diagonalized
on this basis (at leading order in the large-$N_c$ expansion).  We note, however, 
that a recent study by Minahan \cite{Minahan:2005jq} indicates that mixing might occur in the
$\su(2)$ sector in the strong coupling regime.}
There is a separate spin chain Hamiltonian $H_{\su(2)}^{(2n)}$ 
appropriate for each order in the $\lambda$ expansion:
\be
H_{\su(2)} = N + \sum_n g^{2n} H_{\su(2)}^{(2n)}\ ,
\ee
where $g^2 \equiv \lambda/8\pi^2$.  The loop expansion determines 
the generic form of the $n^{\rm th}$-order Hamiltonian, since 
$H_{\su(2)}^{(2n)}$ contains interactions that exchange fields 
separated by at most $n$ lattice sites.
At one-loop order, $H_{\su(2)}^{(2)}$ is given explicitly by
\be
H_{\su(2)}^{(2)} = 2 \sum_{k=1}^N (1-P_{k,k+1})\ ,
\ee
where $P_{k,k+1}$ exchanges fields on the $k^{\rm th}$ and 
$(k+1)^{\rm th}$ lattice sites.  
(The two- and three-loop extensions are given in 
\cite{Beisert:2003tq,Beisert:2003ys,Beisert:2004ry}.)
The eigenvalues of this Hamiltonian can be obtained 
by solving the following Bethe equation
\be
\label{bethe1}
e^{i p_k L} =  \prod_{j=1\atop j\neq k}^N S(p_k,p_j)\ ,
\ee
where the closed-chain $\su(2)$ scattering matrix $S(p_k,p_j)$ is given by
\be
S(p_k,p_j) = \frac{u_k - u_j +i}{u_k - u_j - i}\ .
\ee
The product on the right hand side of eqn.~(\ref{bethe1}) runs over all 
impurity excitations, excluding the $k^{\rm th}$ impurity, and
the Bethe roots $u_k$ encode the pseudomomenta $p_k$ of lattice excitations
via the relation 
\be
u_k = \frac{1}{2}\cot \frac{p_k}{2}\ .
\ee

The form of (\ref{bethe1}) can be understood
intuitively: the exponent on the left-hand-side represents the
phase acquired by a pseudoparticle excitation as it is transported
once around the lattice, and the equation states that this phase must
be equal to that which is acquired by scattering the excitation 
off of every other impurity as is passes around the chain.
The $S$ matrix interpretation of such equations was 
recently studied in detail in \cite{Staudacher:2004tk}, 
where it was emphasized that the form of scattering matrices is
tightly restricted by the constraints of integrability 
(scattering terms must be factorizable into pieces that encode at most
two-body interactions;  
for related results relying on a virial expansion of the multi-loop
spin chain Hamiltonians in ${\cal N}=4$ SYM theory, see \cite{Callan:2004dt}).
For gauge-traced operators, which map to spin chain systems with periodic boundary
conditions (i.e.,~closed spin chains), eqn.~(\ref{bethe1}) is supplemented
by the condition
\be
\label{bethe2}
1 = \prod_{k=1}^N \left( \frac{u_k + i/2}{u_k-i/2}\right)\ ,
\ee
which enforces momentum conservation on the lattice and 
corresponds to the usual level matching condition in the dual sector of closed 
IIB string states in $AdS_5\times S^5$.  (In other words, the set of mode numbers 
$\{n_k\}$ of physical states on the lattice are required by eqn.~(\ref{bethe2})
to satisfy $\sum_k n_k = 0$.)

The Hamiltonian of this system appears as one in an infinite tower of
conserved charges in the theory which, as noted above, is typical of integrable systems.
The entire infinite set of charges is diagonalized by the Bethe ansatz
in eqns.~(\ref{bethe1}, \ref{bethe2}),
and the eigenvalues associated with each charge can be expressed 
in terms of the Bethe roots $u_k$. Operator anomalous dimensions
at $O(\lambda)$ in this sector of the gauge theory 
are thus identified with the energy eigenvalues of the $\su(2)$
one-loop spin chain:
\be
E_{\su(2)}(\{n_i\}) = \frac{\lambda}{8\pi^2} \sum_{k=1}^N \frac{1}{u_k^2 + 1/4}\ .
\ee
The task of computing anomalous dimensions in this sector
of ${\cal N}=4$ SYM theory is thereby reduced to that of solving 
eqns.~(\ref{bethe1}, \ref{bethe2}) for the Bethe roots $u_k$.

This application of the Bethe ansatz
was extended to the full theory at one-loop order by
Beisert and Staudacher in \cite{Beisert:2003yb}, where the authors formulated a set 
of Bethe equations for the full set of $\psu(2,2|4)$ fields 
in the theory.  This methodology was extended beyond one-loop order in $\lambda$
in \cite{Serban:2004jf}, where Serban and Staudacher used an Inozemtsev spin chain 
\cite{Inozemtsev:1989yq,Inozemtsev:2002vb} to capture 
higher-loop effects.  (Such systems are referred to 
as long-range spin chains, because the Hamiltonian can mix spins that
are not on neighboring lattice sites.)  The energies (anomalous dimensions) 
admitted by the Inozemtsev system, however, do not conform to the requirements of
BMN scaling at all orders in perturbation theory
(i.e., at $O(\lambda^n)$, anomalous dimensions should scale at finite $R$-charge 
as $1/R^{2n}$ or, in the language of the string angular momentum, as $1/J^{2n}$).
Beisert, Dippel and Staudacher were subsequently able to formulate a long-range
Bethe ansatz that met the constraints of BMN scaling and correctly 
reproduced $\su(2)$ anomalous dimensions to three-loop order 
in $\lambda$ \cite{Beisert:2004hm}.

Higher-loop physics in the $\su(2)$ sector of the gauge theory is captured by the
following generalized $S$ matrix:
\be
S_{\su(2)}(p_k,p_j) = \frac{\phi(p_k) - \phi(p_j) +i}{\phi(p_k) - \phi(p_j) -i}\ ,
\label{smatrix1}
\ee
where the functions $\phi(p_k)$ are given by
\be
\phi(p_k)\equiv \frac{1}{2}\cot \left(\frac{p_k}{2}\right) 
	\sqrt{1+8 g^2 \sin^2\left(\frac{p_k}{2}\right)}\ .
\ee
(We note, however, that this scattering matrix is derived asymptotically and,
strictly speaking, is only valid for long spin chains.)
Multi-loop operator dimensions are identified with the energies
\be
\label{chainenergy}
E_{\su(2)}(\{n_i\},N,R) = L+g^2 \sum_{k=1}^{N} q_2(p_k)\ ,
\ee
where the functions $q_r(p_k)$ are defined as
\be
q_r(p_k)\equiv \frac{1}{g^{r-1}}\frac{2 \sin\left(\frac{p_k}{2}(r-1)\right)}{r-1}
        \left(\frac{\sqrt{1+8 g^2 \sin^2\left(\frac{p_k}{2}\right)} -1}
	{2g\sin\left(
        \frac{p_k}{2}\right)}\right)^{r-1}\ .
\ee

As described in the introduction, these formulas lead to a general
disagreement with string theory at three-loop order in $\lambda$ and
beyond.  The anomalous dimensions implied by the above scattering matrix disagree
in this respect for all physical $N$-impurity states with the corresponding near-pp-wave energy spectrum 
of the closed string theory on $AdS_5\times S^5$ 
\cite{Callan:2003xr,Callan:2004uv,Callan:2004ev,McLoughlin:2004dh}.
In a remarkable paper by Arutyunov, Frolov and Staudacher \cite{Arutyunov:2004vx}, 
the authors discovered a modification to the long-range gauge 
theory $S$ matrix in eqn.~(\ref{smatrix1}) 
that yields predictions consistent with the near-pp-wave energy spectrum of
the string theory, and exhibits the correct $\lambda^{1/4}$ 
scaling behavior at strong coupling \cite{Gubser:1998bc}.
They accomplished this by including an additional phase contribution $\Phi(p_k,p_j)$ to the
$\su(2)$ scattering matrix in eqn.~(\ref{smatrix1}).  
The resulting {\it quantum string} scattering matrix $S^{\rm IIB}_{S^5}$
takes the form
\be
S_{S^5}^{\rm IIB}(p_k,p_j) = S_{\su(2)}(p_k,p_j) \Phi(p_k,p_j)\ ,
\label{IIBsmatrixsu2}
\ee
where $\Phi(p_k,p_j)$ is given by 
\be
\label{phi}
\Phi(p_k,p_j) \equiv \exp\left[ 2i \sum_{r=0}^\infty
	\left(\frac{g^2}{2}\right)^{r+2}
	\left(q_{r+2}(p_k)q_{r+3}(p_j) -q_{r+3}(p_k)q_{r+2}(p_j) \right)
	\right]\ .
\ee
It was demonstrated 
in \cite{McLoughlin:2004dh} that the energies implied by the corresponding 
quantum string Bethe equations 
perfectly match energy spectra derived directly from the string theory for
general closed string states in the $\su(2)$ sector (in the near-pp-wave limit).  
Since the energy spectra contain rather specialized structures 
(as we have already seen in the case of the open string states in 
section~\ref{strings}), this is a non-trivial result:  it stands as evidence 
that the quantum string theory is integrable, at least in this limit.  

In what follows we apply similar methods to the $\su(2)$ and $\Sl(2)$ sectors
of the bosonic open string states studied above.  Inspired by the Bethe ansatz 
methodology presented at one-loop order for the matching of open string energies
with the dimensions of corresponding gauge theory operators in 
\cite{Chen:2004mu,Berenstein:2002zw,DeWolfe:2004zt,Lee:2002cu},
we derive long-range Bethe equations that diagonalize the interacting
Hamiltonian $H_{\rm BB}$ on completely general open string states.

\subsection{Dirichlet $SO(4)_{S^5}$ ($\su(2)$) ansatz}
On the open spin chain, pseudoparticle excitations (or magnons) pass from one end
of the chain, scatter off of the boundary and return to the 
starting point.  Each transition across the length of the chain
contributes a factor of $\exp( ip  L )$ to the phase 
(the effect on the phase should be additive, so the momentum is 
understood to be oriented in the positive direction
after it scatters from the boundary).  The total phase acquired in this
process must be equal to that generated by scattering from 
each impurity on the chain as it travels to the boundary and back 
(after scattering from the boundary, the 
relative momenta of the other impurities on the chain acquire an overall
sign change).  We use this simple picture to propose a long-range 
Bethe ansatz for the $\su(2)$ Dirichlet sector of bosonic open string
states:\footnote{This ansatz is motivated by gauge theory proposals at one-loop order
in $\lambda$ \cite{Chen:2004mu,DeWolfe:2004zt}.  We note, however, that the specific
equation given in \cite{DeWolfe:2004zt} appears with a sign discrepancy.}
\be
e^{2i p_k L} &=&  \prod_{j=1\atop j\neq k}^N S_{S^5}^{\rm IIB}(p_k,p_j)
		S_{S^5}^{\rm IIB}(p_k,-p_j)
\nn\\
	&=&  \prod_{j=1\atop j\neq k}^N 
	S_{\su(2)}(p_k,p_j) S_{\su(2)}(p_k,-p_j) 
	\Phi(p_k,p_j)\Phi(p_k,-p_j)\ .
\label{BAstring}
\ee
The long-range scattering matrices $S_{\su(2)}$ are 
defined in eqn.~(\ref{smatrix1}) above, and the phase corrections $\Phi(p_k,p_j)$ 
for the string are given in eqn.~(\ref{phi}).
We might have included some prefactor dependent on $p_k$
to capture phase contributions due to scattering
from the boundary itself.  
We will see, however, that in this sector such contributions turn out to be trivial 
(this situation changes when we examine the $\Sl(2)$ sector below).
We also note that for open spin chains there is no momentum condition that 
enforces level matching among the mode numbers (just as there is no level
matching condition for the open string states studied in section~\ref{strings}).

To simplify the presentation, and to review a general strategy for 
solving eqn.~(\ref{BAstring}) \cite{Minahan:2002ve,Arutyunov:2004vx}, 
we will first compute energy spectra in the near-pp-wave
limit for length-$L$ states of the open chain whose impurities do not form bound states.
These are the bosonic, symmetric-traceless open string states 
given in eqn.~(\ref{s5states}) above, with $(n_1 \neq n_2 \neq \cdots \neq n_N)$. 
We must solve the Bethe ansatz in eqn.~(\ref{BAstring}) order by order
in the large-$J$ expansion and extract contributions to the energy spectrum at $O(1/J)$.
In terms of the string angular momentum $J$ and the total impurity number $N$, 
the lattice length $L$ for the corresponding integrable spin chain is
\be
\label{Lsu2}
L = J-1+N\ .
\ee
When we analyze the dual gauge theory in section~\ref{SYMBA} below we will see 
how this equation is determined.  For the moment, however, we will proceed
without further motivation.

A useful technique, as demonstrated in \cite{Minahan:2002ve} and \cite{Arutyunov:2004vx}, 
for example, is to expand the excitation momenta according to
\be
\label{pexpand}
p_k = \frac{\pi n_k}{J} +\frac{p_k^{(1)}}{J^{3/2}} + \frac{p_k^{(2)}}{J^2} + O(1/J^{5/2})\ ,
\ee
and solve eqn.~(\ref{BAstring}) for $p_k^{(1)}$,  $p_k^{(2)}$, etc.
The leading contribution carries the integer mode number $n_k$ of the pseudomomentum
$p_k$.
(To compute energy eigenvalues to $O(1/J)$, we will not
need to find contributions to $p_k$ beyond $O(1/J^2)$.)
We first note that the coefficient $p_k^{(1)}$ is nonzero only in the presence of
bound momentum states, where some set of $p_k$ share the same mode number
$n_k$.   In the absence of bound states, the contribution to eqn.~(\ref{BAstring}) from
the phase factors $\Phi(p_k,p_j)$ is
\be
\label{phi1}
\Phi(p_k,p_j) \Phi(p_k,-p_j) &=& 1 - \frac{2 i n_k \pi}{J p_- (n_k^2-n_j^2)}
	\left[ n_k^2(\omega_{n_j} - p_-) - n_j^2(\omega_{n_k} - p_-) \right]
	+ O(1/J^2)\ .
\nn\\
&&
\ee
The Bethe equation in (\ref{BAstring}) is then satisfied by 
\be
p_k^{(2)} = -n_k\pi (N-1) - \sum_{j=1\atop j\neq k}^N
	\frac{ n_k \pi}{ p_- (n_k^2-n_j^2)}
	\left[ n_k^2(\omega_{n_j} - p_-) + n_j^2(\omega_{n_k} + p_-) \right]\ .
\ee
From eqn.~(\ref{chainenergy}) above, we therefore find the following 
contribution to the energy spectrum at $O(1/J)$:
\be
\delta E(\{n_i\},N,J) 
	= \frac{1}{J}\sum_{k=1}^N \frac{ n_k p_k^{(2)} }{4 \pi p_- \omega_{n_k} }
	= \frac{1}{J}\sum_{j,k =1 \atop j\neq k}^N
	\frac{n_k^2 + n_j^2 \lambda' \omega_{n_k}^2}{\omega_{n_j}\omega_{n_k}}\ .
\ee
This matches the corresponding energy correction for string states with no
overlapping mode numbers computed directly from the string lightcone Hamiltonian
in eqn.~(\ref{nonconfluentE}).

Solving the Bethe equation in the presence of bound states is slightly
more complicated.  
To align with conventions in the literature we will adopt a notation similar to that
presented in \cite{Arutyunov:2004vx}, 
wherein the complete set of $N$ mode numbers associated
with each impurity excitation is divided into $M$ subsets of equal mode numbers
labeled by the index $k$:
\be
\bigl\{  \underset{k=1}{\underbrace{\{n_1,n_1,\ldots,n_1\}}},  
	\underset{k=2}{\underbrace{\{n_2,n_2,\ldots,n_2\}}}, \ldots , 
	\underset{k=M}{\underbrace{\{n_M,n_M,\ldots,n_M\}}}
\bigr\}\ .
\label{subsets}
\ee
The $k^{\rm th}$ subset contains $N_k$ total 
numbers $n_k$, which are individually labeled by an index $m_k \in 1,\ldots,N_k$.
The Bethe ansatz therefore takes the general form
\be
\exp (2 i  p_{k,m_k} L)  &=& 
	\prod_{{l_k}=1 \atop {l_k} \neq {m_k}}^{N_k} 
	S_{S^5}^{\rm IIB}(p_{k,m_k},p_{k,l_k})S_{S^5}^{\rm IIB}(p_{k,m_k},-p_{k,l_k})
\nn\\
&&\kern-00pt
	\times
	\prod_{j=1\atop j\neq k}^M \prod_{{m_j}=1}^{N_j}
	S_{S^5}^{\rm IIB}(p_{k,m_k},p_{j,m_j})S_{S^5}^{\rm IIB}(p_{k,m_k},-p_{j,m_j})\ ,
\label{BAstring2}
\ee
where the momenta now carry the labels $k$ and $m_k$, and 
are expanded in large $J$ according to 
\be
\label{pexp}
p_{k,m_k}&=&\frac{\pi n_k}{J}+\frac{p^{(1)}_{k,m_k}}{J^{3/2}}+\frac{p^{(2)}_{k,m_k}}{J^2}+
	O(1/J^{5/2})\ .
\ee
Within each bound state, the contribution from factors of the phase correction  
$\Phi(p_{k,m_k},p_{k,l_k})$ is
\be
\label{phi2}
\Phi(p_{k,m_k},p_{k,l_k}) \Phi(p_{k,m_k},-p_{k,l_k}) = 1-\frac{i n_k\pi }{4 p_-\omega_{n_k} J}
	\left(8p_-^2+n_k^2-8p_-\omega_{n_k} \right) 
	+ O(1/J^2)\ ,
\ee
and the coefficient $p_{k,m_k}^{(1)}$ is now nonzero in the presence of bound states:
\be
\label{p1}
p_{k,m_k}^{(1)}= - \frac{n_k^2 \pi^2 \omega_{n_k}}{p_-} 
	\sum_{l_k=1 \atop l_k\neq m_k}^{N_k} \frac{1}{p_{k,m_k}^{(1)}-p_{k,l_k}^{(1)}}\ .
\ee
As for the closed-chain case in \cite{Arutyunov:2004vx}, we have found a generalized 
Stieltjes problem \cite{Shastry:2001} which is solved by setting  
\be
\left( p_{k,m_k}^{(1)}\right)^2 = -\frac{n_k^2 \pi^2 \omega_{n_k}}{2p_-} u_{N_k,m_k}^2\ ,
\label{ST1}
\ee
where $u_{N_k,m_k}$ are the roots of the Hermite polynomials $Q_{N_k}(u)$ satisfying 
$Q''(u)-u Q'(u)+ N_k Q(u)=0$.

At the next subleading order we find that $p_{k,m_k}^{(2)}$ is given by  
\be
p_{k,m_k}^{(2)} &=& (1-N)\pi n_k
	-\frac{1}{2}\sum_{l_k=1 \atop l_k\neq m_k}^{N_k}
	\left[ p_{k,m_k}^{(2)} - p_{k,l_k}^{(2)} 
	+n_k\pi\left(\frac{\omega_{n_k}}{p_-}-2\right)\right]
\nn\\
&&	-\sum_{j=1\atop j\neq k}^M \frac{n_k \pi N_j}{p_-(n_k^2-n_j^2)}
	\left[ n_k^2(\omega_{n_j} - p_-) + n_j^2(\omega_{n_k} + p_-) \right]\ .
\label{p2string}
\ee
We can simplify matters by noting that when we expand the energy formula
in eqn.~(\ref{chainenergy}) to the order of interest, the correction at $O(1/J)$ 
involves a sum over the index $m_k$:
\be
E(\{n_k\}) 
	&=& J + \sum_{k=1}^M \frac{2 N_k \omega_{n_k} }{p_-}
	+ \frac{ 1 }{p_- J}
	\sum_{k=1}^M \sum_{m_k=1}^{N_k}
	\frac{1}{8\pi^2 \omega_{n_k}^3}
	\left[ p_-^2 \left(p_{k,m_k}^{(1)}\right)^2
	+ 2\pi n_k \omega_{n_k}^2 p_{k,m_k}^{(2)} \right]\ .
\nn\\
\label{Eexp}
\ee
Under this sum the roots $u_{N_k,m_k}$ of the Hermite polynomials $Q_{N_k}(u)$ satisfy
\be
\sum_{m_k=1}^{N_k} u_{N_k,m_k}^2 = N_k(N_k-1)\ .
\ee
Furthermore, contributions to the right hand side of eqn.~(\ref{p2string})
involving $p_{k,m_k}^{(2)}$ and $p_{k,l_k}^{(2)}$ will cancel:
\be
\sum_{m_k=1}^{N_k} \frac{n_k}{4\pi\omega_{n_k}}p_{k,m_k}^{(2)} = 
	\frac{n_k^2}{8p_-}N_k(1-N_k)
	+\sum_{j=1\atop j\neq k}^N \frac{N_j N_k}{8p_- (n_j^2-n_k^2)}
	\left( n_k^4\frac{\omega_{n_j}}{\omega_{n_k}} - 
	n_j^4 \frac{\omega_{n_k}}{\omega_{n_j}}
	\right)\ .
\ee
We therefore arrive at a final expression for the correction to the 
energy spectrum at $O(1/J)$ in the near-pp-wave limit:
\be
\delta E_{S^5}(\{n_i\},\{N_{i}\},M,J)  &=&  -\frac{1}{8 J}\biggl\{
	\sum_{k=1}^M N_{k}(N_{k}-1) \frac{n_k^2(6 +n_k^2\lambda')}{4 \omega_{n_k}^2}
\nn\\
&&\kern+40pt
	+ \sum_{j,k =1 \atop j\neq k}^M N_{j} N_{k}
	\frac{n_k^2 + n_j^2 \lambda' \omega_{n_k}^2}{\omega_{n_j}\omega_{n_k}}
	\biggr\}\ .
\ee
This matches the general $\su(2)$ string theory prediction in eqn.~(\ref{confluentE}).

\subsection{Neumann $SO(4)_{AdS}$ ($\Sl(2)$) ansatz}
We can extend this analysis to the 
protected $\Sl(2)$ sector of bosonic open string states.  
The open-string Bethe equation we propose is based on the long-range
quantum string ansatz presented in \cite{Staudacher:2004tk} for the 
corresponding sector of $\Sl(2)$ closed-string states:
\be
e^{i p_k L} =  \prod_{j=1\atop j\neq k}^N S_{AdS}^{\rm IIB}(p_k,p_j)\ ,
\ee 
where the scattering matrix $S_{AdS}^{\rm IIB}$ is given by
\be
\label{betheSL2}
S_{AdS}^{\rm IIB}(p_{k},p_{j}) \equiv
	\frac{\phi(p_{k}) - \phi(p_{j}) - i}{\phi(p_{k}) - \phi(p_{j}) + i}\,
	\Psi(p_{k},p_{p_j})\ ,
\label{IIBsmatrixsl2}
\ee
and the phase $\Psi(p_{k},p_{p_j})$ is defined as
\be
\Psi(p_k,p_j) \equiv \exp\left[ - 2i \sum_{r=0}^\infty
	\left(\frac{g^2}{2}\right)^{r+1}
	\left(q_{r+1}(p_k)q_{r+2}(p_j) -q_{r+2}(p_k)q_{r+1}(p_j) \right)
	\right]\ .
\ee
This ansatz was derived in \cite{Staudacher:2004tk} using general considerations
of integrability and of the structure of the near-pp-wave energy spectrum 
computed from the string theory \cite{McLoughlin:2004dh}.  
The lattice length $L$ obeys
\be
\label{Lsl2}
L = J-1\ .
\ee
This formula is again based on gauge theory considerations, and will be discussed in 
section~\ref{SYMBA} below.

Using this scattering matrix, we propose the following general Bethe ansatz 
(allowing for confluent mode numbers) for the $\Sl(2)$ sector of open string states
fluctuating in the Neumann directions of the $AdS_5$ subspace:
\be
\exp (2 i  p_{k,m_k} L)  &=& 
	e^{-2 i p_{k,m_k}}\prod_{{l_k}=1 \atop {l_k} \neq {m_k}}^{N_k} 
	S_{AdS}^{\rm IIB}(p_{k,m_k},p_{k,l_k})S_{AdS}^{\rm IIB}(p_{k,m_k},-p_{k,l_k})
\nn\\
&&\kern-0pt
	\times
	\prod_{j=1\atop j\neq k}^M \prod_{{m_j}=1}^{N_j}
	S_{AdS}^{\rm IIB}(p_{k,m_k},p_{j,m_j})S_{AdS}^{\rm IIB}(p_{k,m_k},-p_{j,m_j})\ .
\label{SL2ansatz}
\ee
The factor $e^{-2 i p_{k,m_k}}$ on the right-hand side is a phase contribution interpreted as being
associated with scattering from the boundary.
Note also that in the small-$\lambda'$ expansion contributions to the Bethe equation 
from $\Psi(p_k,p_j)$ enter at a lower order than those for $\Phi(p_k,p_j)$ in the 
$\su(2)$ case considered above.

Expanding the pseudoparticle momenta according to eqn.~(\ref{pexp}), we find
the following equation at first subleading order:
\be
p_{k,m_k}^{(1)} = \frac{n_k^2 \pi^2 \omega_{n_k}}{p_-}
	\sum_{l_k = 1\atop l_k\neq m_k}^{N_k}
	\frac{1}{p_{k,m_k}^{(1)}-p_{k,l_k}^{(1)}}\ .
\ee
This differs by an overall sign 
from the corresponding $\su(2)$ equation computed in (\ref{p1}) above.  
At next order in the $1/J$ expansion we find
\be
p_{k,m_k}^{(2)} = \frac{1}{2}\sum_{l_k=1\atop l_k\neq m_k}^{N_k}
	\left( p_{k,l_k}^{(2)} - p_{k,m_k}^{(2)} - \frac{n_k \pi}{p_-}\omega_{n_k}\right)
	+ \sum_{j=1\atop j\neq k}^M N_j \frac{F_1(n_j,n_k)}{F_2(n_j,n_k)}\ ,
\ee
where, for convenience, we have defined
\be
F_1(n_j,n_k) & \equiv & 
	- n_j^2 n_k^2 \pi \biggl\{
	8 + (n_j^2+n_k^2)\lambda' 
\nn\\
&&
\kern-40pt
	- \sqrt{\lambda'}\Bigl[
	(8+n_k(n_k-n_j)\lambda')\omega_{n_j}
	+ \omega_{n_k}(8+n_j(n_j-n_k)\lambda' - 8\sqrt{\lambda'} \omega_{n_j})
	\Bigr] \biggr\}\ ,
\\
F_2(n_j,n_k) & \equiv & 
	(n_j^2-n_k^2)\Bigl[
	-4 + n_j n_k\lambda' 
	+ 4\sqrt{\lambda'}(\omega_{n_k} +\omega_{n_j}-\sqrt{\lambda'}\omega_{n_k}\omega_{n_j})
	\Bigr]\ .
\ee
Using the general expansion of the energy in eqn.~(\ref{Eexp}) above, we 
arrive at the following formula for the $O(1/J)$ energy shift in the
near-pp-wave limit:
\be
\delta E_{AdS}(\{n_i\},\{N_i\},M,J)  &=&  
	-\frac{1}{32\,J}\biggl\{ 
	\sum_{k=1}^M N_k (N_k-1) \frac{n_k^2 (2+n_k^2\lambda')}{\omega_{n_k}^2}
	+\sum_{j,k=1 \atop j\neq k}^M N_j N_k 
	\frac{n_j^2 n_k^2 \lambda'}{\omega_{n_j}\omega_{n_k}}
	\biggr\}\ .
\nn\\
&&
\ee
This is precisely the energy expression computed for the $\Sl(2)$ 
sector of the string theory in eqn.~(\ref{EAdS}) above.

\section{SYM long-range Bethe equations}
\label{SYMBA}
As noted in section~\ref{StringBE}, 
a long-range Bethe ansatz for the protected sector of single-trace
$\su(2)$ operators in ${\cal N}=4$ SYM theory was formulated in \cite{Beisert:2004hm},
and an analogous $\Sl(2)$ ansatz was given more recently in \cite{Staudacher:2004tk}.
Building on our successful derivation of the open-string $\su(2)$ 
Bethe equations in (\ref{BAstring}, \ref{BAstring2}) based on corresponding equations 
formulated for closed string states, we can immediately write down 
long-range, open-chain 
Bethe ansatz equations for $\su(2)$ and $\Sl(2)$ sectors of the ${\cal N}=2$ defect conformal
field theory dual to the corresponding open string states described in section~\ref{strings}.  
The theory we are interested in 
describes a stack of $N_c$ coincident $D3$-branes with one $D5$-brane extended
along three of the worldvolume dimensions of the $D3$ stack (see table~\ref{dims}).
This theory, studied in \cite{DeWolfe:2004zt,Lee:2002cu,DeWolfe:2001pq}, 
contains a three-dimensional ${\cal N}=4$ $SU(N_c)$ 
hypermultiplet in addition to the bulk four-dimensional hypermultiplet.
The $D5$-brane defect preserves an $SO(3,2)$ subgroup of the conformal group 
and eight of the supersymmetries, though it breaks the $R$-symmetry from $SU(4)$ to 
$SU(2)_H\times SU(2)_V$.  We can decompose the bulk ${\cal N}=4$ 
vector multiplet into a three-dimensional ${\cal N}=4$ vector multiplet and a 
three-dimensional ${\cal N}=4$ adjoint hypermultiplet. 
The vector multiplet contains the following bosonic fields:
\be
A_{\mu}\ , \qquad X^1\ , \qquad X^2\ ,\qquad  X^3\ ,\qquad  D_3 X^I\ ,
\nn
\ee
with $\mu \in 0,1,2$ and $I\in 4,5,6$.  The hypermultiplet contains the 
component of the gauge field normal to the defect, as well
as the scalars $X^4$, $X^5$, $X^6$ and $D_3 X^A$, with $A\in 1,2,3$. 
The three-dimensional ${\cal N}=4$ $SU(N_c)$ hypermultiplet contains 
a set of complex scalars $q^m$ (with $m\in 1,2$) 
which couple canonically to the gauge fields. 
On the string side we take the Penrose limit by boosting in the
direction dual to the $(1,2)$ plane. 
The charge $J^3_{SU(2)_H}$ is identified with the string angular momentum $J$
(which we will refer to simply as the $R$-charge), and 
the fields of interest are thus charged under $SU(2)_H$ according to
\be
\begin{array}{ccc}
X^1,~X^2,~X^3~~ (J=1)~, \qquad X^4,~X^5,~X^6~~ (J=0)~,
\qquad q^m,~\tilde{q}^m~~ (J=1/2)\ .
\nn
\end{array}
\ee
The Bethe ground state is taken to be
\be
\bar{q_1} Z \cdots Z q_2\ ,
\nn
\ee
with $Z\equiv X^1+iX^2$.  
(The indices on the fields $q^m$ are lowered by Chan-Paton factors; see 
\cite{DeWolfe:2004zt} for further details.)
$R$-charge impurities in the $\su(2)$ sector 
are denoted by $W\equiv X^4+iX^5$, 
and we form a basis of length-$L$ (i.e.,~operator monomials containing a total of $L$ fields), 
$N$-impurity operators given by
\be
\bar{q_1} W^N Z^{L-N-2} q_2\ , \qquad 
\bar{q_1} W^{N-1} Z W Z^{L-N-3} q_2\ , \qquad 
\bar{q_1} W^{N-2} Z W^2 Z^{L-N-4} q_2\ , \qquad \ldots
\label{su2sector}
\ee
This basis corresponds to the $\su(2)$ sector of open Dirichlet string states
in the $S^5$ subspace.
With the $R$-charge assignments given above, we now have the relation
given in eqn.~(\ref{Lsu2}):
\be
L_{\su(2)} = J-1+N\ .
\ee
The sector of $\Sl(2)$ open string states with Neumann boundary conditions is dual to
operators with derivative insertions (with $D \equiv D_1 + iD_2$): 
\be
\bar{q_1} D^N Z^{L-N-2} q_2\ , \qquad 
\bar{q_1} D^{N-1} Z D Z^{L-N-3} q_2\ , \qquad 
\bar{q_1} D^{N-2} Z D^2 Z^{L-N-4} q_2\ , \qquad \ldots
\label{sl2sector}
\ee
so that the lattice length of the
corresponding $\Sl(2)$ open spin chain is 
\be
L_{\Sl(2)}=J-1\ ,
\ee 
as given in eqn.~(\ref{Lsl2}).
By mapping the dilatation 
generator to the Hamiltonian of an integrable open spin chain,  
Bethe ansatz equations for similar sectors of operators were derived at one-loop order
in $\lambda$ in \cite{DeWolfe:2004zt,Chen:2004mu}.

\subsection{Long-range $\su(2)$ ansatz}
We will focus first on the closed-chain, long-range scattering matrix for the
$\su(2)$ spin chain, derived in \cite{Beisert:2004hm} and given above in 
eqn.~(\ref{smatrix1}).  In terms of this $S$ matrix, 
we propose the following long-range, open-chain 
ansatz for this sector:
\be
e^{2ip_k L}= \prod_{j\not= k}^{N} S_{\su(2)}(p_k,p_j) S_{\su(2)}(p_k,-p_j)\ .
\label{SYMBA1}
\ee
As discussed in section~\ref{StringBE}, 
we could include a phase contribution due to scattering at the boundaries.
Given what we have learned from the corresponding $\su(2)$ string Bethe equations, 
however, we expect that such contributions are trivial.

The full Bethe equation takes the form
\be
\exp (2 i L p_{k,m_k})  &=& 
	\prod_{{l_k}=1 \atop {l_k} \neq {m_k}}^{N_k} 
	S_{\su(2)}(p_{k,m_k},p_{k,l_k})S_{\su(2)}(p_{k,m_k},-p_{k,l_k})
\nn\\
&&	\times
	\prod_{j=1\atop j\neq k}^M \prod_{{m_j}=1}^{N_j}
	S_{\su(2)}(p_{k,m_k},p_{j,m_j})S_{\su(2)}(p_{k,m_k},-p_{j,m_j})\ .
\label{BAstring3}
\ee
Expanding pseudoparticle momenta in the usual fashion
\be
p_{k,m_k} = \frac{n_k \pi}{J} + \frac{p_{k,m_k}^{(1)}}{J^{3/2}} + \frac{p_{k,m_k}^{(2)}}{J} + 
	O(1/J^{3/2})\ ,
\ee
we find that eqn.~(\ref{BAstring3}) is satisfied to first subleading order within 
the $k^{\rm th}$ subset of overlapping mode numbers for
\be
p_{k,m_k}^{(1)} = -\frac{n_k^2\pi^2 \omega_{n_k}}{p_-}\sum_{l_k = 1 \atop l_k\neq m_k}^{N_k}
	\frac{1}{p_{k,m_k}^{(1)}-p_{k,l_k}^{(1)}}\ .
\ee
As in the case of the string Bethe ansatz above, this Stieltjes problem 
is solved by setting
\be
\left( p_{k,m_k}^{(1)}\right)^2 = -\frac{\pi^2 n_k^2 \omega_{n_k}}{2p_-} u_{N_k,m_k}^2\ ,
\label{ST2}
\ee
which is identical to the string theory solution in eqn.~(\ref{ST1}) above.

At the next subleading order we find
\be
p_{k,m_k}^{(2)}  =  (1-N)\pi n_k 
	- \frac{1}{2}\sum_{l_k=1\atop l_k\neq m_k}^{N_k}\left(
	p_{k,m_k}^{(2)} - p_{k,l_k}^{(2)} - \frac{p_- \pi n_k}{\omega_{n_k}} \right)
	+ \sum_{j=1\atop j\neq k}^M N_j\frac{2\pi n_k n_j^2 \omega_{n_k}}{p_-(n_j^2-n_k^2)}\ .
\ee
The contribution from $p_{k,m_k}^{(2)}$ to the $O(1/J)$ energy 
correction again involves a sum over the index $m_k$,
so that contributions to $\delta E$ from terms proportional to 
$p_{k,m_k}^{(2)} - p_{k,l_k}^{(2)}$ will cancel.  
Proceeding in the same fashion as before, 
we find a final expression for the energy spectrum at
$O(1/J)$ in the near-BMN limit:
\be
\delta E_{\su(2)}(\{n_i\},\{N_i\},M,J) 
	&=& \frac{1}{p_- J}
	\sum_{k=1}^M \frac{n_k^2}{16~\omega_{n_k}^2 }
		N_k(N_k-1)(p_- - 4~\omega_{n_k})
\nn\\
&&	- \frac{1}{p_- J}
	\sum_{k,j=1\atop j\neq k}^M N_j N_k  
	 \frac{n_k^2}{4~\omega_{n_k}}\ .
\ee 
The energy formula again breaks neatly into contributions associated with 
scattering within and between bound momentum states of the spin chain.
Expanding in small $\lambda'$, we obtain
\be
\delta E_{\su(2)}(\{n_i\},\{N_i\},M,J)
	&=& 
	\frac{1}{J}\sum_{k=1}^M N_k(N_k-1) 
	\Bigl[ 
	-\frac{3}{16} n_k^2\lambda'
	+ \frac{1}{64}n_k^4(\lambda')^2-\frac{1}{512}n_k^6(\lambda')^3  \Bigr]
\nn\\
&&
	+ \frac{1}{J}\sum_{k,j=1\atop j\neq k}^M N_j N_k 
	\Bigl[
	-\frac{1}{8}(n_j^2+n_k^2)\lambda' + \frac{1}{64}(n_j^4+n_k^4)(\lambda')^2
\nn\\
&&
\kern+80pt
	-\frac{3}{1024}(n_j^6+n_k^6)(\lambda')^3 \Bigr]
	 +O((\lambda')^4)\ .
\ee
Once again, we find perfect agreement with the corresponding $\su(2)$ string theory
predictions in eqn.~(\ref{stringEXP}) at one- and two-loop order, but
a disagreement at three loops and beyond!

\subsection{Long-range $\Sl(2)$ ansatz}
One might guess that the correct closed-chain ${\cal N}=4$
$\Sl(2)$ scattering matrix can be obtained from eqn.~(\ref{betheSL2}) simply by 
eliminating the phase contribution from $\Psi(p_k,p_j)$.  From the structure
of the phase factor $\Psi(p_k,p_j)$, we can immediately see that this would lead
to a general mismatch at two-loop order and beyond.
It turns out that this simple hypothesis is incorrect.  The correct asymptotic 
gauge theory scattering matrix was derived in \cite{Staudacher:2004tk}:
\be
S_{\Sl(2)}(p_k,p_j) \equiv 
	\frac{\phi(p_{k}) - \phi(p_{j}) - i}{\phi(p_{k}) - \phi(p_{j}) + i}
	e^{2i\theta(p_k,p_j)}\ .
\label{sl2matrix}
\ee
The phase shift $\theta(p_k,p_j)$ was originally formulated in the computation 
of a long-range scattering matrix for the fermionic $\su(1|1)$ sector of ${\cal N}=4$ SYM
via Staudacher's method of the perturbative asymptotic Bethe ansatz
(see \cite{Staudacher:2004tk} for further details):
\be
S_{\su(1|1)}(p_k,p_j) = -e^{i\theta(p_k,p_j)}\ .
\ee
The $\Sl(2)$ $S$ matrix in eqn.~(\ref{sl2matrix})
was then derived from this $\su(1|1)$ scattering matrix and the corresponding matrix 
in the $\su(2)$ sector (\ref{smatrix1}) using the following remarkable relationship:
\be
S_{\Sl(2)} = S_{\su(1|1)}S^{-1}_{\su(2)}S_{\su(1|1)}\ .
\ee
According to \cite{Staudacher:2004tk}, this equation appears to hold for both 
the string theory and gauge theory.
The phase $\theta(p_k,p_j)$ was calculated perturbatively in 
small $\lambda$ in \cite{Staudacher:2004tk}, yielding:
\be
\theta(p_k,p_j) & = & 
	\Bigl[
	2g^2\sin^2\frac{p_k}{2} \sin p_j 
	-2 g^4 \sin^4 \frac{p_k}{2} \sin 2p_j
	+ 8 g^4 \sin p_k \sin^2 \frac{p_k}{2} \sin^2 \frac{p_j}{2}
	\Bigr]
\nn\\
&&
	- (p_k \leftrightharpoons p_j)\ .
\label{SL2phase}
\ee
The expansion to this order ($O(\lambda^2)$) makes the resulting scattering matrix
(\ref{sl2matrix}) accurate up to and including three-loop 
order in $\lambda$; a higher-order expansion would require specific knowledge of 
certain four-loop vertices in the ${\cal N}=4$ 
gauge theory.\footnote{We thank Matthias Staudacher for clarification on this point.}

In terms of the $\Sl(2)$ $S$ matrix in eqn.~(\ref{sl2matrix}), we arrive at the following
long-range, open-chain Bethe ansatz for the $\Sl(2)$ sector (\ref{sl2sector}) of the 
defect conformal field theory:
\be
\exp(2i p_k L) = e^{-2i p_k} \prod_{j=1 \atop j\neq k}^N 
	S_{\Sl(2)}(p_k,p_j)S_{\Sl(2)}(p_k,-p_j)\ .
\ee
The additional phase shift $e^{-2i p_k}$ includes the effects of boundary scattering,
analogous to the corresponding open string ansatz in eqn.~(\ref{SL2ansatz}).
We can solve this equation in the near-BMN limit using the methods 
described in detail above. 
Omitting the computational details, we find the following general energy shift
(for all possible mode-index distributions) in the near-BMN limit:
\be
\delta E_{\Sl(2)}(\{n_i\},\{N_i\},M,J)
	&=& 
	\frac{1}{J}\sum_{k=1}^M N_k(1-N_k) \Bigl[
	\frac{1}{16}n_k^2 \lambda'
	+\frac{1}{64}n_k^4 (\lambda')^2
	-  \frac{3}{512}n_k^6 (\lambda')^3 
	\Bigr]
\nn\\
&&\kern-00pt
	+\frac{1}{J}\sum_{j,k=1\atop j\neq k}^M N_j N_k 
	\Bigl[ -\frac{1}{32} n_j^2 n_k^2 (\lambda')^2 
	+ \frac{5}{1024}n_j^2 n_k^2 (n_j^2 + n_k^2)(\lambda')^3
	\Bigr]
\nn\\
&&	+ O((\lambda')^4)\ .
\ee
Comparing with the corresponding expansion in the near-pp-wave limit
of the open-string $\Sl(2)$ Neumann sector in eqn.~(\ref{SL2EXP}),
we again find precise agreement at one- and two-loop order in 
$\lambda'$, followed by a mismatch at three loops and beyond.

\section{Summary and discussion}
We have formulated a set of long-range 
Bethe ans\"atze for open string states that arise from standard deformations of
IIB string theory on $AdS_5\times S^5$.\footnote{Let us note that 
our methods may also be useful in more general deformations of the theory, 
some examples of which were considered recently in \cite{Frolov:2005ty}.}
In a bosonic symmetric-traceless $S^5$ sector we studied open string
states with Dirichlet boundary conditions, labeled by an $\su(2)$ subalgebra of
$\psu(2,2|4)$.  
The open-string Bethe ansatz is formulated in this sector 
from the closed-string scattering matrix $S^{\rm IIB}_{S^5}(p_k,p_j)$ in 
eqn.~(\ref{IIBsmatrixsu2}) according to 
\be
&&{\rm closed\ \su(2)\ strings:} \qquad e^{i p_k L} = \prod_{j\neq k} S^{\rm IIB}_{S^5}(p_k,p_j)\ , \\
&&{\rm open\ \su(2)\ strings:} \qquad e^{2 i p_k L} = \prod_{j\neq k} 
			S^{\rm IIB}_{S^5}(p_k,p_j)S^{\rm IIB}_{S^5}(p_k,-p_j)\ .
\label{S5ansatz}
\ee
In this sector we found that there is no phase contribution
from boundary scattering. 
The corresponding prescription in the sector of $\Sl(2)$ 
symmetric-traceless open string states fluctuating in $AdS_5$ with
Neumann boundary conditions was found to be
\be
&&{\rm closed\ \Sl(2)\ strings:} \qquad e^{i p_k L} = \prod_{j\neq k} S^{\rm IIB}_{AdS}(p_k,p_j)\ , \\
&&{\rm open\ \Sl(2)\ strings:} \qquad e^{2 i p_k L} = e^{-2 i p_{k}}\prod_{j\neq k} 
			S^{\rm IIB}_{AdS}(p_k,p_j)S^{\rm IIB}_{AdS}(p_k,-p_j)\ .
\label{AdSansatz}
\ee
The closed-string scattering matrix $S^{\rm IIB}_{AdS}(p_k,p_j)$ appears in 
eqn.~(\ref{IIBsmatrixsl2}) above.  In this sector the open-string Bethe equation
receives a contribution that may be interpreted as being 
due to boundary scattering; this appears as the
$e^{-2 i p_{k}}$ prefactor on the right-hand side of eqn.~(\ref{AdSansatz}).
In both sectors we computed the near-pp-wave energy spectra of general $N$-impurity 
states directly from the string theory.  The Bethe ans\"atze presented here 
reproduce these results precisely, which is consistent with the expectations of
integrability.

When the above prescription is applied to the corresponding 
sectors of operators in the dual ${\cal N}=2$ defect conformal field theory, we obtain
\be
\label{bethestringsu2}
&&{\rm closed\ \su(2)\ SYM\ spin\ chains:} \qquad e^{i p_k L} 
	= \prod_{j\neq k} S_{\su(2)}(p_k,p_j)\ , \\
\label{bethechainsu2}
&&{\rm open\ \su(2)\ SYM\ spin\ chains:} \qquad e^{2 i p_k L} = \prod_{j\neq k} 
			S_{\su(2)}(p_k,p_j)S_{\su(2)}(p_k,-p_j)\ ,
\ee
in the $\su(2)$ sector, and
\be
\label{bethestringsl2}
\kern-25pt
&&{\rm closed\ \Sl(2)\ SYM\ spin\ chains:} \qquad e^{i p_k L} 
	= \prod_{j\neq k} S_{\Sl(2)}(p_k,p_j)\ , \\
\label{bethechainsl2}
\kern-25pt
&&{\rm open\ \Sl(2)\ SYM\ spin\ chains:} \qquad e^{2 i p_k L} = 
			e^{-2 i p_{k}}\prod_{j\neq k} 
			S_{\Sl(2)}(p_k,p_j)S_{\Sl(2)}(p_k,-p_j)\ ,
\ee
in the $\Sl(2)$ sector.
We find that the operator anomalous dimensions (or energy spectra of the corresponding
spin chains) in both the ${\cal N}=4$ (\ref{bethestringsu2}, \ref{bethestringsl2}) and 
${\cal N}=2$ (\ref{bethechainsu2}, \ref{bethechainsl2}) gauge theories
match the energy spectra of general $\su(2)$ and $\Sl(2)$ physical 
string states to one- and two-loop order in $\lambda'$, 
but disagree at three-loop order and higher.

\begin{figure}[htb]
\begin{center}
\includegraphics[width=3.5in,height=.8in,angle=0]{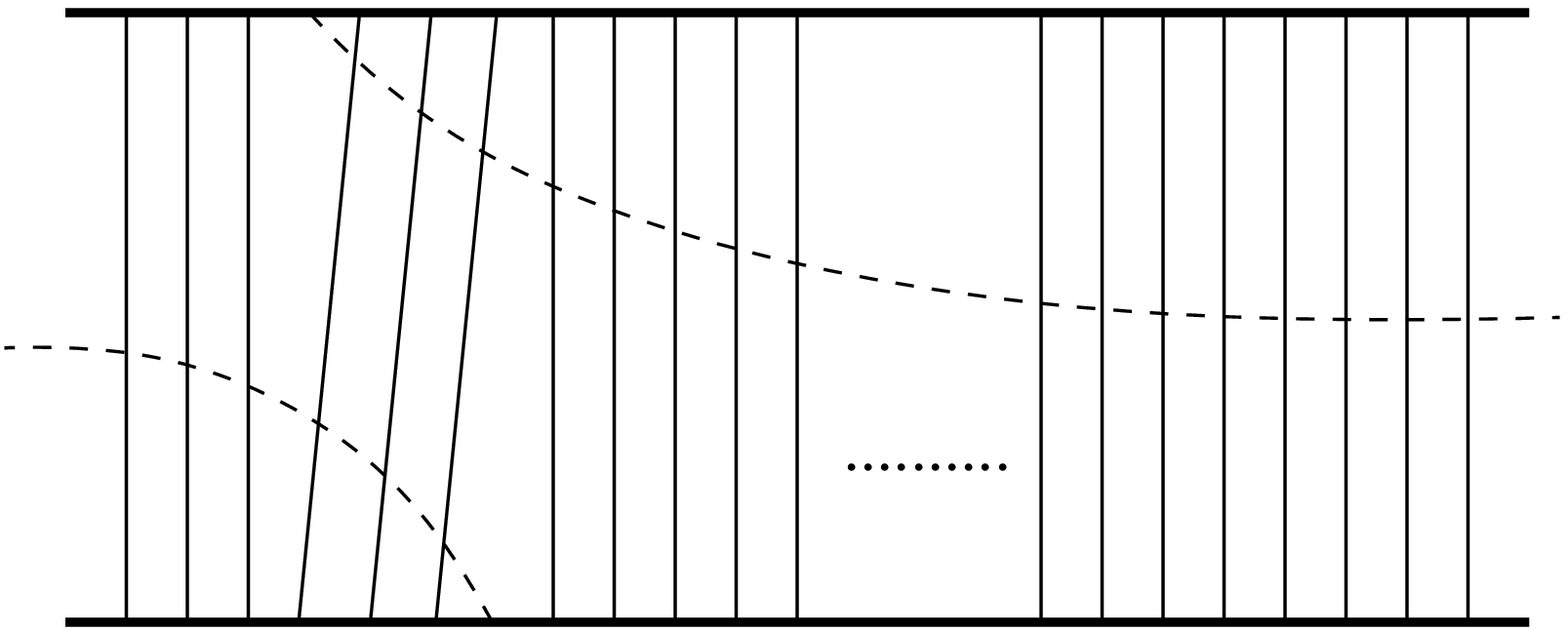}
\caption{Wrapping interaction diagram}
\label{wrapfig}
\end{center}
\end{figure}

\begin{figure}[htb]
\begin{center}
\includegraphics[width=3.5in,height=.8in,angle=0]{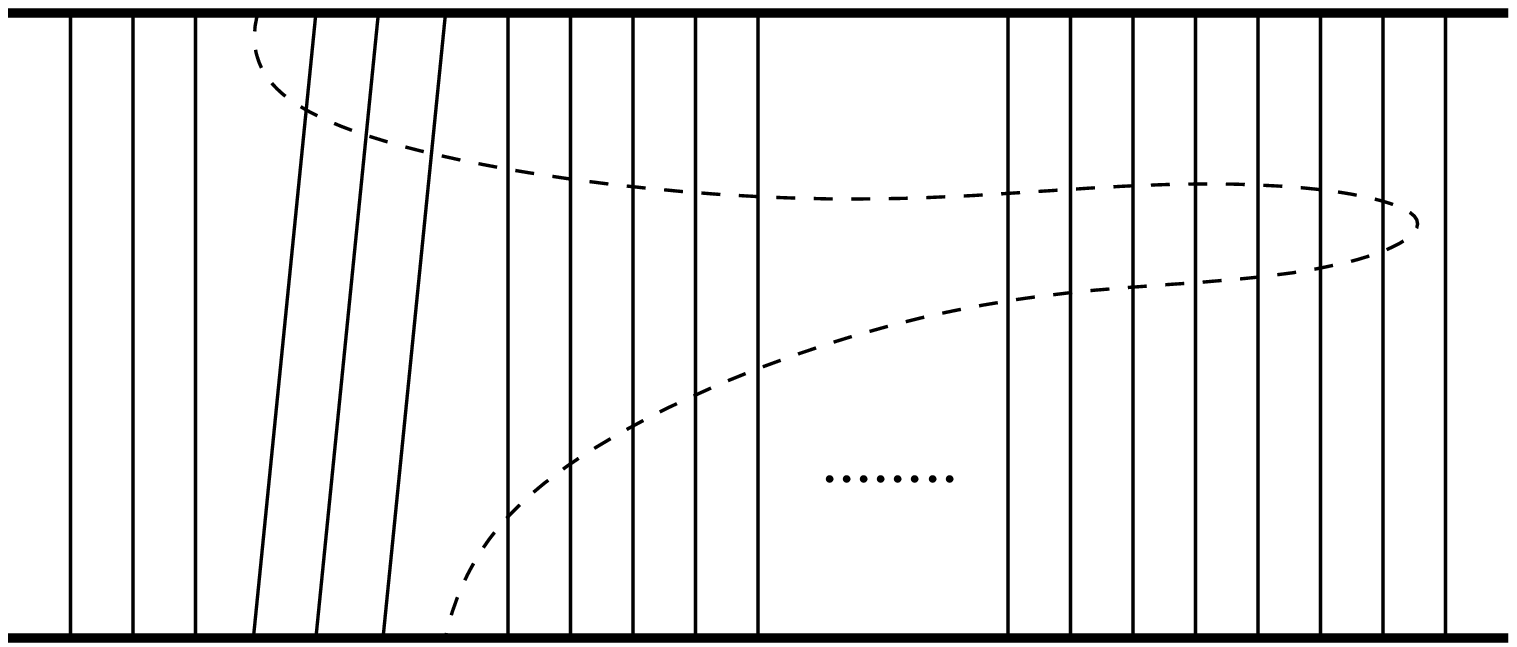}
\caption{A long-range, non-wrapping diagram}
\label{longfig}
\end{center}
\end{figure}

In \cite{Beisert:2004hm}, Beisert, Dippel and Staudacher suggested
a generic type of term that could lead to order-of-limits issues
in the comparison of string and gauge theory in this 
setting:\footnote{Apart from the fact that such terms
might explain the mismatch with string theory beyond two loops, there is no
other reason to suspect that the gauge theory produces such functions.  See
\cite{Beisert:2004hm} for details.}
\be
\frac{\lambda^L}{(c+\lambda)^L}\ . \nn
\ee 
Terms of this form will lead to different contributions depending on the 
sequence of limits taken in figure~\ref{fig1} above.
While the range of interactions associated with such terms would 
typically be greater than the length
of the chain, they would not necessarily have to wrap {\it around} the spin chain: 
they need only involve a number of vertices that exceeds the total number 
of lattice sites on the chain.  For closed chains, this could obviously 
include wrapping interactions, pictured schematically in figure~\ref{wrapfig}. 
For both open and closed chains, however, these interactions might also be of 
the form shown in figure~\ref{longfig},
which does not require periodicity on the lattice.  
We note that, by definition, both types of diagram are
absent in the limit of infinite chain length.  (Other examples
include interactions that are defined to stretch between the
boundaries of an open chain, or contain a large number of vertices
associated with other intermediate states.)

In the $\su(2)$ sector, 
the difference between the closed-chain scattering matrices for the string
and gauge theory is collected into the phase $\Phi(p_k,p_j)$:
\be
S^{\rm IIB}_{S^5}(p_k,p_j) =  S_{\su(2)}(p_k,p_j)\Phi(p_k,p_j)\ .
\ee
If the general disagreement between string and gauge theory is
due to wrapping interactions alone, we should expect that
the contribution from this phase correction to the open-string Bethe ansatz
\be
\Phi(p_k,p_j)\Phi(p_k,-p_j) 
\label{openphi}
\ee
will drop out.  While some terms from $\Phi(p_k,p_j)$ do cancel in
eqn.~(\ref{bethechainsu2}), there is still a nonzero contribution 
from these phases to the open string energy spectrum (see eqns.~(\ref{phi1})
and (\ref{phi2})).  
One conjecture is that this observation in fact isolates stringy contributions 
to the Bethe equations from each type of interaction pictured in figures~\ref{wrapfig} and
\ref{longfig}. 
Contributions from long-range wrapping (figure~\ref{wrapfig}) 
and non-wrapping (figure~\ref{longfig}) interactions
might then be encoded by those terms in $\Phi(p_k,p_j)$
that do or do not cancel in the phase correction (\ref{openphi}) to the open-string
Bethe equations, respectively.

Another interpretation of our results is that the prescription for 
extending the closed-string, long-range Bethe ansatz to open strings 
(\ref{S5ansatz}, \ref{AdSansatz}) cannot
be carried over analogously to the dual gauge theory (\ref{bethechainsu2}, \ref{bethechainsl2}).  
It is a logical possibility that the discrepancy between the closed-string
and closed-chain energy spectra is due solely to wrapping interactions, and that in 
moving to the open-chain formulation of the gauge theory we have eliminated all order-of-limits
issues, but at the same time introduced some other set of 
unwanted contributions.  While this seems unlikely, a direct higher-loop 
gauge theory calculation might settle any uncertainty.

\section*{Acknowledgments}
We would like to thank John Schwarz and Curt Callan for their ongoing support
and encouragement.  We also thank Niklas Beisert, Andrei Mikhailov 
and Matthias Staudacher for explanations and useful discussions.
This work was supported by US Department of Energy 
grant DE-FG03-92-ER40701.

\bibliographystyle{utcaps}
\bibliography{Open}

\end{document}